\documentclass[twocolumn]{article}

\usepackage{amsmath, amsthm, amsfonts}
\usepackage{hyperref}
\usepackage{natbib}
\usepackage{graphicx}
\usepackage{geometry}

\theoremstyle{definition}

\theoremstyle{remark}

\title{Large-scale asymmetry in galaxy spin directions - analysis of galaxies with spectra in DES, SDSS, and DESI Legacy Survey}

\author{Lior Shamir \\
\small Kansas State University \\
\small Manhattan, KS 66506, USA
}

\date{}

\newgeometry{vmargin={20mm}, hmargin={17mm,17mm}} 

\begin{document}
\maketitle

\begin{abstract}

Multiple previous studies using several different probes have shown considerable evidence for the existence of cosmological-scale anisotropy and a Hubble-scale axis. One of the probes that show such evidence is the distribution of the directions toward which galaxies spin. The advantage of the analysis of the distribution of galaxy spin directions compared to the CMB anisotropy is that the ratio of galaxy spin directions is a relative measurement, and therefore less sensitive to background contamination such as Milky Way obstruction. Another advantage is that many spiral galaxies have spectra, and therefore allow to analyze the location of such axis relative to Earth. This paper shows an analysis of the distribution of the spin directions of over 90K galaxies with spectra. That analysis is also compared to previous analyses using the Earth-based SDSS, Pan-STARRS, and DESI Legacy Survey, as well as space-based data collected by HST. The results show very good agreement between the distribution patterns observed with the different telescopes. The dipole or quadrupole axes formed by the spin directions of the galaxies with spectra do not necessarily go directly through Earth.

\end{abstract}

\section{Introduction}
\label{introduction}

Recent observations with several different probes have shown accumulating evidence of cosmological-scale anisotropy, and the presence of a Hubble-scale axis. Perhaps the most notable and thoroughly studied probe showing evidence of cosmological-scale anisotropy and a cosmological-scale axis is the cosmic microwave background \citep{eriksen2004asymmetries,cline2003does,gordon2004low,campanelli2007cosmic,zhe2015quadrupole,abramo2006anomalies,mariano2013cmb,land2005examination,ade2014planck,santos2015influence,dong2015inflation,gruppuso2018evens,yeung2022directional}. Other messengers that show cosmological anisotropy and possible axes in the large-scale structure include radio sources \citep{ghosh2016probing,tiwari2015dipole,tiwari2016revisiting}, LX-T scaling \citep{migkas2020probing}, short gamma ray bursts \citep{meszaros2019oppositeness}, cosmological acceleration rates \citep{perivolaropoulos2014large,migkas2021cosmological,krishnan2021hints}, galaxy morphology types \citep{javanmardi2017anisotropy}, Ia supernova \citep{javanmardi2015probing,lin2016significance}, dark energy \citep{adhav2011kantowski,adhav2011lrs,perivolaropoulos2014large,colin2019evidence}, fine structure constant \citep{webb2011indications}, galaxy motion \citep{skeivalas2021predictive}, $H_o$ \citep{luongo2021larger}, polarization of quasars \citep{hutsemekers2005mapping,secrest2021test,zhao2021tomographic,semenaite2021cosmological}, and high-energy cosmic rays \citep{aab2017observation}. 

Other studies showed that the large-scale distribution of galaxies in the Universe is not random, as discussed by \cite{de1986slice,hawkins20032df,colless20032df,jones2005scaling,deng2006super,adelman2008ma} and others. Data-driven observations based on large datasets revealed the existence of very large structures \citep{gott2005map,lietzen2016discovery,horvath2015new} that could be beyond astrophysical scale, and therefore challenge the cosmological principle.


These observations can be viewed as a certain tension with the standard cosmological models \citep{pecker1997some,perivolaropoulos2014large,bull2016beyond,velten2020hubble,krishnan2021hints,luongo2021larger}, and triggered several expansions to the standard models, as well as other cosmological theories that shift from the standard models.  Possible explanations and theories include double inflation \citep{feng2003double}, primordial anisotropic vacuum pressure \citep{rodrigues2008anisotropic}, contraction prior to inflation \citep{piao2004suppressing}, flat space cosmology \citep{tatum2018flat,tatum2018clues,azarnia2021islands}, multiple vacua \citep{piao2005possible}, spinor-driven inflation \citep{bohmer2008cmb}, and moving dark energy \citep{jimenez2007cosmology}. 

Other proposed theories can be related to the geometry of the Universe such as ellipsoidal universe \citep{campanelli2006ellipsoidal,campanelli2007cosmic,campanelli2011cosmic,gruppuso2007complete,cea2014ellipsoidal}, geometric inflation \citep{arciniega2020geometric,edelstein2020aspects,arciniega2020towards,jaime2021viability}, supersymmetric flows \citep{rajpoot2017supersymmetric}, and rotating universe \citep{godel1949example}. Early rotating universe theories were based on a non-expanding universe \citep{godel1949example}, and therefore conflict with the observation that the Universe is expanding. More recent models of rotating Universe were modified to support cosmological expansion  \citep{ozsvath1962finite,ozsvath2001approaches,sivaram2012primordial,chechin2016rotation,seshavatharam2020integrated,camp2021}.

The existence of a cosmological-scale axis can also be associated with the theory of black hole cosmology, and can explain cosmic accelerated inflation without the assumption of dark energy \citep{pathria1972universe,stuckey1994observable,easson2001universe,chakrabarty2020toy}. Black holes spin  \citep{gammie2004black,takahashi2004shapes,volonteri2005distribution,mcclintock2006spin,mudambi2020estimation,reynolds2021observational}, and their spin is inherited from the spin of the star from which the black hole was created \citep{mcclintock2006spin}. Due to the spin of the black hole, it has been proposed that a universe hosted in a black hole should have an axis and a preferred direction \citep{poplawski2010cosmology,seshavatharam2010physics,seshavatharam2014understanding,christillin2014machian,seshavatharam2020light,seshavatharam2020integrated}. Black hole cosmology is also associated with the theory of holographic universe \citep{susskind1995world,bak2000holographic,bousso2002holographic,myung2005holographic,hu2006interacting,rinaldi2022matrix}, which can also represent the large-scale structure of the Universe in a hierarchical manner \citep{sivaram2013holography,shor2021representation}.

In addition to the messengers discussed above, multiple previous studies showed substantial evidence that the distribution of the spin directions of spiral galaxies is anisotropic, and forms a cosmological-scale axis \citep{macgillivray1985anisotropy,longo2011detection,shamir2012handedness,shamir2013color,shamir2016asymmetry,shamir2017photometric,shamir2017large,shamir2017colour,shamir2019large,shamir2020pasa,shamir2020patterns,shamir2020large,shamir2020asymmetry,lee2019galaxy,lee2019mysterious,shamir2021particles,shamir2021large,shamir2022new}. These observations include different telescopes such as SDSS \citep{shamir2012handedness,shamir2020patterns,shamir2021particles,shamir2022new}, Pan-STARRS \citep{shamir2020patterns}, HST \citep{shamir2020pasa}, and DECam \citep{shamir2021large,shamir2022new}. These telescopes show consistent patterns of the asymmetry, regardless of the telescope being used or the method of annotation of the galaxies \citep{shamir2021large,shamir2022new}.

The alignment of the spin directions of galaxies was observed within cosmic web filaments, as discussed in \citep{tempel2013evidence,tempel2013galaxy,tempel2014detecting,dubois2014dancing,kraljic2021sdss} among other studies, but also between galaxies too far from each other to interact gravitationally \citep{lee2019mysterious}. That alignment is difficult to explain with the standard gravity models, and was defined as ``mysterious'' \citep{lee2019mysterious}. It has also been proposed that the galaxy spin direction is a probe for studying the early Universe \citep{motloch2021observed}.

As a relative measurement, the probe of the large-scale distribution of galaxy spin directions has the advantage of being less sensitive to background contamination such as Milky Way obstruction. The reason is that the asymmetry in galaxy spin directions in a certain field is determined by the difference between the number of galaxies spinning clockwise and the number of galaxies spinning counterclockwise in the same field. Since all galaxies are observed in the exact same field, any background contamination or other effect that affects galaxies spinning clockwise is naturally expected to affect galaxies spinning counterclockwise in the same manner. Because the background contamination affects all galaxies in the field, its existence is not expected to affect the asymmetry. That might be different from other probes such as CMB, where the measurement is absolute, and background contamination that affects a certain field can lead to the observation of anisotropy. 

Another advantage of using spin direction asymmetry is that spiral galaxies are very common in the Universe, and are present in a broad redshift range. That important advantage allows to identify not merely the existence of a cosmological-scale axis, but also to determine its approximated location relative to Earth. This paper uses a dataset of $\sim$100K spiral galaxies with spectra to identify the location of the axis formed by the distribution of these galaxies.

\section{Data}
\label{data}

The dataset used in this study is made of galaxies imaged by three different telescopes: the Sloan Digital Sky Survey (SDSS), the Dark Energy Survey (DES), and the Dark Energy Spectroscopic Instrument (DESI) Legacy Survey. That datasets is compared to the distribution of the spin directions of galaxies in datasets used in previous studies.
 
The directions of the curves of the spiral arm of a galaxy is a reliable indication on the spin direction of the galaxy. For instance, \cite{de1958tilt} used  dust silhouette and Doppler shift to determine that in all tested cases the spiral arms were trailing, and therefore allow to determine the spin direction of the galaxy. While in some rare cases galaxies can have leading arms, such as NGC 4622 \citep{freeman1991simulating}, the vast majority of spiral galaxies have trailing arms, and therefore the curve of the galaxy arms can in most cases determine the direction towards which it spins.

The primary task related to the data used in this study is the annotation of galaxies by their spin direction. Although one of the datasets used here was annotated manually \citep{shamir2020pasa}, the scale of databases acquired by modern digital sky surveys is far too large for manual annotation. The practical approach to the annotation of very large datasets of galaxy images is by automatic annotation. It should be mentioned that while pattern recognition, and specifically deep neural networks, have become the common solution to automatic annotation of galaxy images, these approaches might not be suitable for studying subtle cosmological-scale anisotropies \citep{dhar2022systematic}. Pattern recognition, and specifically deep neural networks, are based on complex data-driven rules, and are therefore subjected to subtle biases that are very difficult to identify \citep{carter2020overinterpretation,dhar2021evaluation}, and have also been detected in galaxy images \citep{dhar2022systematic}. Such biases might skew the results. A more thorough discussion about the possible impact of bias in the annotation algorithm is provided in Section~\ref{error}.

\subsection{Automatic annotation of clockwise and counterclockwise galaxies}
\label{ganalyzer}

The galaxies were annotated automatically from the 2D galaxy images by using the {\it Ganalyzer} algorithm \citep{shamir2011ganalyzer}. Ganalyzer is a model-driven algorithm that uses fully symmetric clear mathematical rules to determine the spin direction of a spiral galaxy. The algorithm is described in detail in \citep{shamir2011ganalyzer}, and brief description is also available in \citep{dojcsak2014quantitative,hoehn2014characteristics,shamir2017photometric,shamir2017large,shamir2020patterns,shamir2021particles,shamir2021large,shamir2022new}. 

In summary, Ganalyzer first transforms each galaxy image into its radial intensity plot. The radial intensity plot transformation of a galaxy image is a 35$\times$360 image, such that the pixel $(x,y)$ in the radial intensity plot is the median value of the 5$\times$5 pixels around coordinates $(O_x+\sin(\theta) \cdot r,O_y-\cos(\theta)\cdot r)$ in the original galaxy image, where {\it r} is the radial distance in percentage of the galaxy radius, $\theta$ is the polar angle measured in degrees, and $(O_x,O_y)$ are the pixel coordinates of the galaxy center.

Pixels on the galaxy arms are expected to be brighter than pixels that are not on the galaxy arm at the same radial distance from the center. Therefore, peaks in the radial intensity plot are expected to correspond to pixels on the arms of the galaxy at different radial distances from the center. To identify the arms, a peak detection algorithm \citep{morhavc2000identification} is applied to the lines in the radial intensity plot.

\begin{figure}[h]
\centering
\includegraphics[scale=0.44]{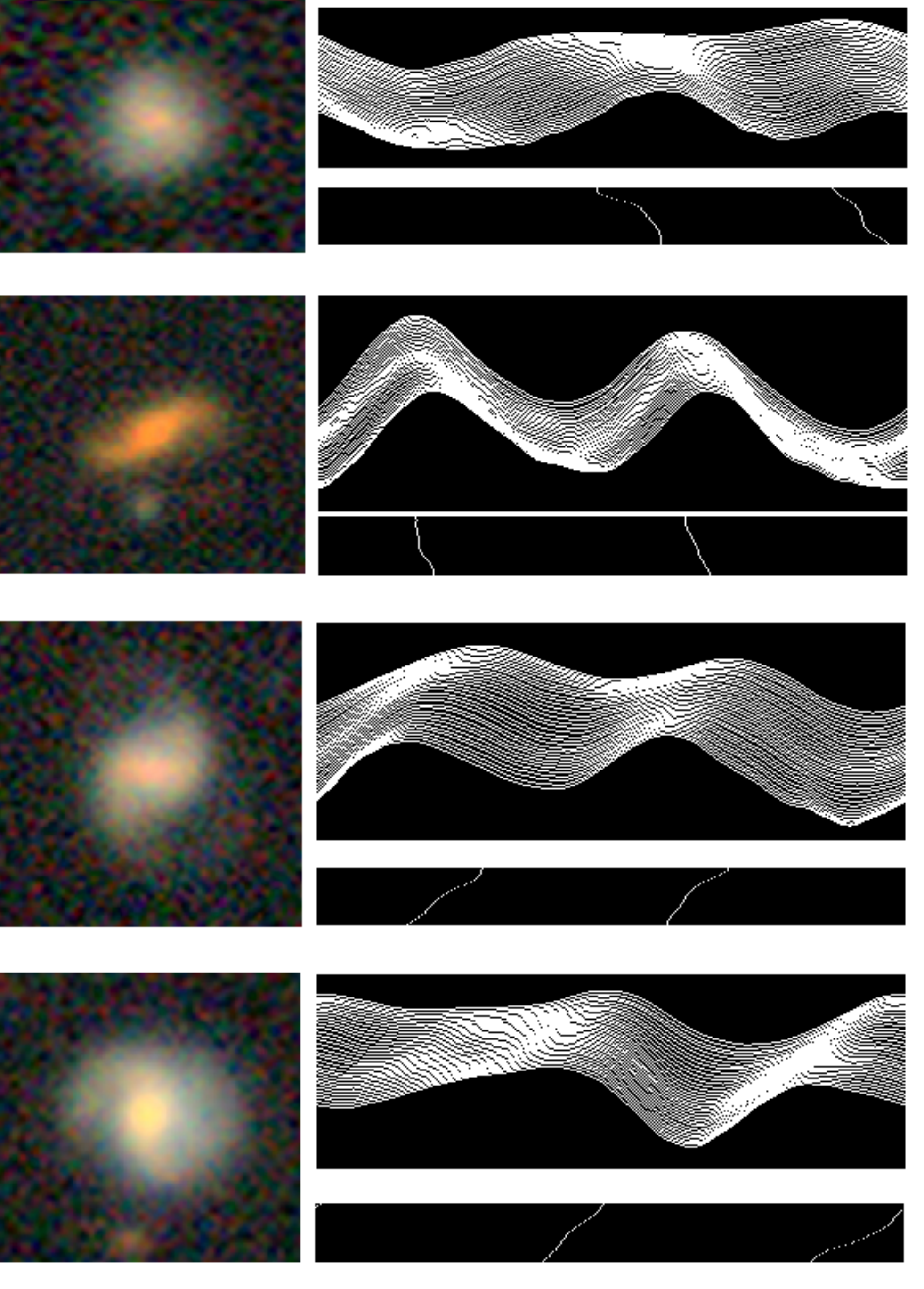}
\caption{Examples of galaxy images and their corresponding radial intensity plots. The peaks detected in the radial intensity plots are displayed below the radial intensity plots. The directions of the lines formed by the peaks reflect the curves of the arms, and consequently the spin direction of the galaxy. The algorithm is symmetric, and it is model-driven with intuitive rules. }
\label{radial_intensity_plots}
\end{figure}

Figure~\ref{radial_intensity_plots} shows examples of the radial intensity plots and the peaks detected in them in four DES galaxies. The figure also shows the radial intensity plot of each galaxy, and the lines formed by the detected peaks. Each line in the radial intensity plot shows the brightness of the pixels around the center of the galaxy. That is, the first value in the line is the brightness of the pixel at 0$^o$ compared to the galaxy center, and the last value in the line is the brightness of the pixel at 359$^o$. Since each line is at a different radius from the center, each radial intensity plot has multiple lines. Below each radial intensity plot shown in Figure~\ref{radial_intensity_plots}, the figure displays the peaks identified in the lines. That is, if a pixel in the radial intensity plot is identified as a peak in its line, the corresponding pixel in the image below the radial intensity plot is white. Otherwise, the pixel is black. More information about Ganalyzer can be found in \citep{shamir2011ganalyzer,dojcsak2014quantitative,shamir2017photometric,shamir2017colour,shamir2017large,shamir2019large,shamir2020patterns,shamir2021large,shamir2022new}.

As Figure~\ref{radial_intensity_plots} shows, the lines formed by the peaks identified in the radial intensity plots form lines in different directions. A linear regression is applied to the peaks in adjunct lines formed by the peaks, and the slope of the linear regression reflects the curve of the arm. As  Figure~\ref{radial_intensity_plots} shows, if the galaxy spin clockwise the slope of the regression is positive, while if the galaxy spins counterclockwise the slope is negative. Therefore, the slope of the regression can be used to determine the spin direction of the galaxy.

Naturally, many galaxies are elliptical galaxies, irregular galaxies, or spiral galaxies that do not have an identifiable spin direction. To avoid galaxies that do not have an identifiable spin direction, only galaxies with 30 or more peaks that form curved lines in the radial intensity plots are used. If that criteria is not met, the galaxy is determined to have an unidentifiable spin direction. As mentioned above, the main advantage of the algorithm is that it follows defined and fully symmetric rules. Analysis of different situations when applying the algorithm to populations of galaxies are described in \cite{shamir2021particles}, and also more briefly in Section~\ref{error}.

\subsection{Galaxy images and digital sky surveys}

The dataset of galaxies with spectra used in this study is based on data collected by three different digital sky surveys: SDSS, DES, and the DESI Legacy Survey. The SDSS galaxies are 63,693 galaxies with spectra used in \citep{shamir2019large,shamir2020patterns,shamir2020pasa}. The preparation of that dataset is described in detail in \citep{shamir2020patterns}. The galaxies from the DESI Legacy Survey are the subset of galaxies used by \cite{shamir2021large} that had spectroscopic redshift through the catalog of \cite{zhou2021clustering}. The entire DESI Legacy Survey dataset contained 807,898 galaxies \citep{shamir2021large}, but only 23,715 of these galaxies had spectroscopic redshift through the catalog of \cite{zhou2021clustering}. The \citep{zhou2021clustering} catalog contains mostly the photometric redshift of the DESI Legacy Survey galaxies. As also discussed in Section~\ref{error}, since the inaccuracy of photometric redshift is greater than the expected signal, the photometric redshift cannot be used for this study, and therefore only galaxies that had spectroscopic redshift are used. 

Similarly to the DESI dataset, galaxies from DES data release (DR) 1 were also used. The initial list of DES objects included all objects identified as exponential disks, de Vaucouleurs ${r}^{1/4}$ profiles, or round exponential galaxies, and were brighter than 20.5 magnitude in one or more of the g, r or z bands. That list contained an initial set of 18,869,713 objects. The galaxy images were downloaded using the {\it cutout} API of the DESI Legacy Survey server, which also provide access to DES data. The size of each image was 256$\times$256, and retrieved in the JPEG format. Each image was scaled using the Petrosian radius to ensure that the galaxy fits in the image. The process of downloading the images started on April 25th 2021, and ended about six months later on November 1st 2021. 

Once the image files were downloaded, they were annotated by their spin direction using the Ganalyzer method described above and in \citep{shamir2011ganalyzer,dojcsak2014quantitative,shamir2017photometric,shamir2017colour,shamir2017large,shamir2019large,shamir2020patterns,shamir2021large,shamir2022new}. The annotation of the galaxies lasted 73 days of operation using a single Intel Xeon processor. Then, the images were mirrored using {\it ImageMagick} and annotated again to allow repeating the experiments with mirrored images. That provided a dataset of 773,068 galaxies annotated by their spin directions. To remove satelite galaxies or stars positioned inside a galaxy, objects that had another object in the dataset within 0.01$^o$ or less were removed from the dataset. That provided a dataset of 739,286 galaxies imaged by DES, and 14,365 of these galaxies had redshift through the 2dF redshift survey \citep{colless20032df,cole20052df}.

\begin{figure}[h]
\centering
\includegraphics[scale=0.7]{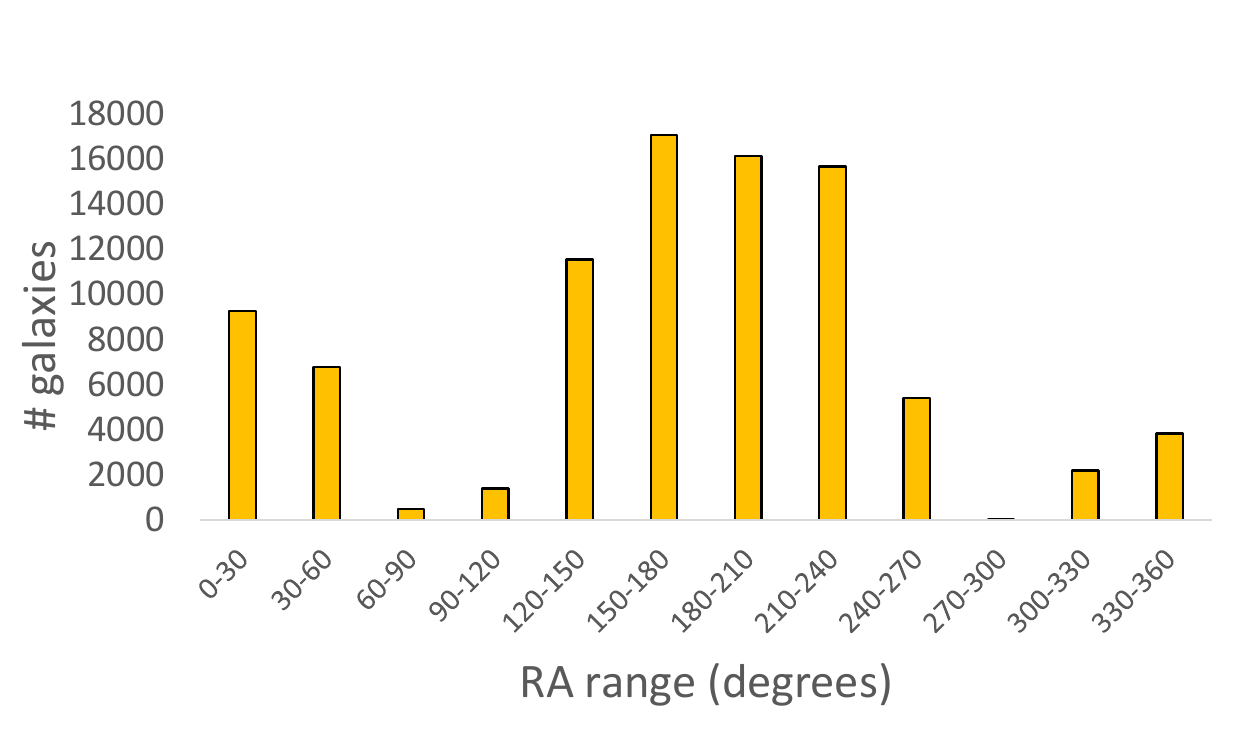}
\caption{The distribution of the galaxies in the dataset in different 30$^o$ RA ranges.}
\label{ra_distribution}
\end{figure}

\begin{figure}[h]
\centering
\includegraphics[scale=0.65]{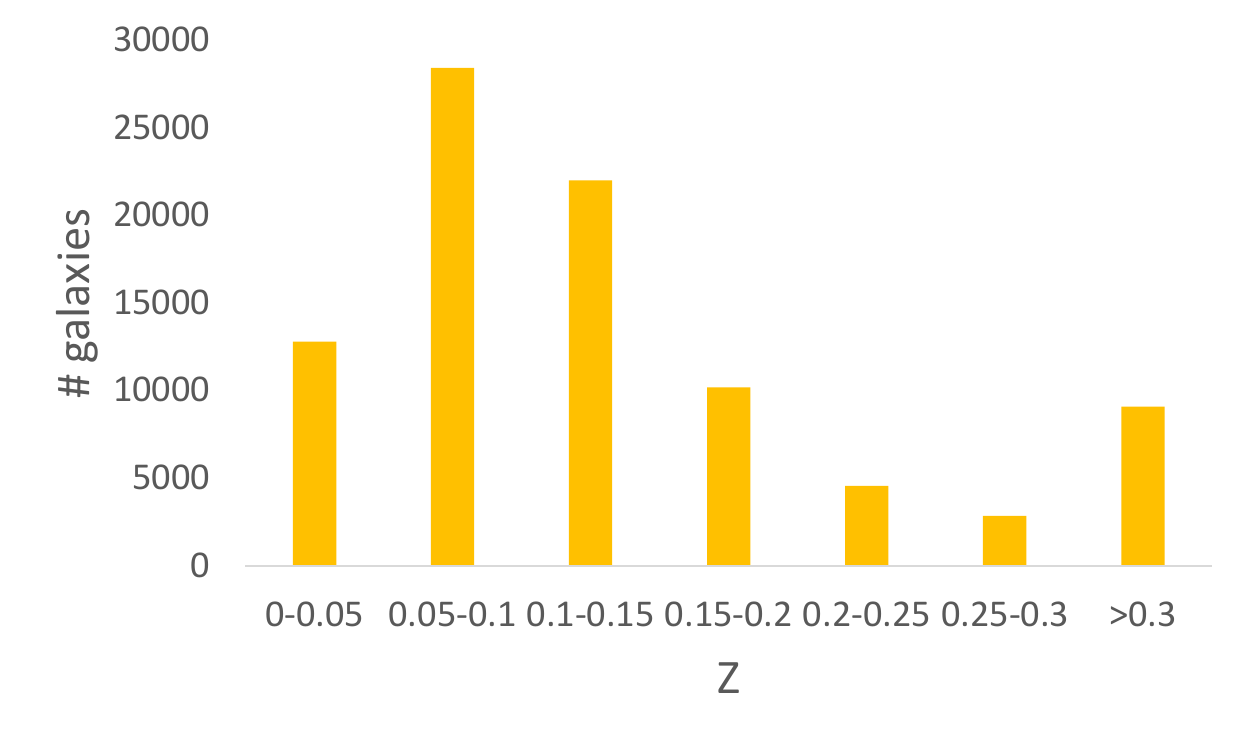}
\caption{The redshift distribution of the galaxies.}
\label{z_distribution}
\end{figure}

\begin{figure}[h]
\centering
\includegraphics[scale=0.15]{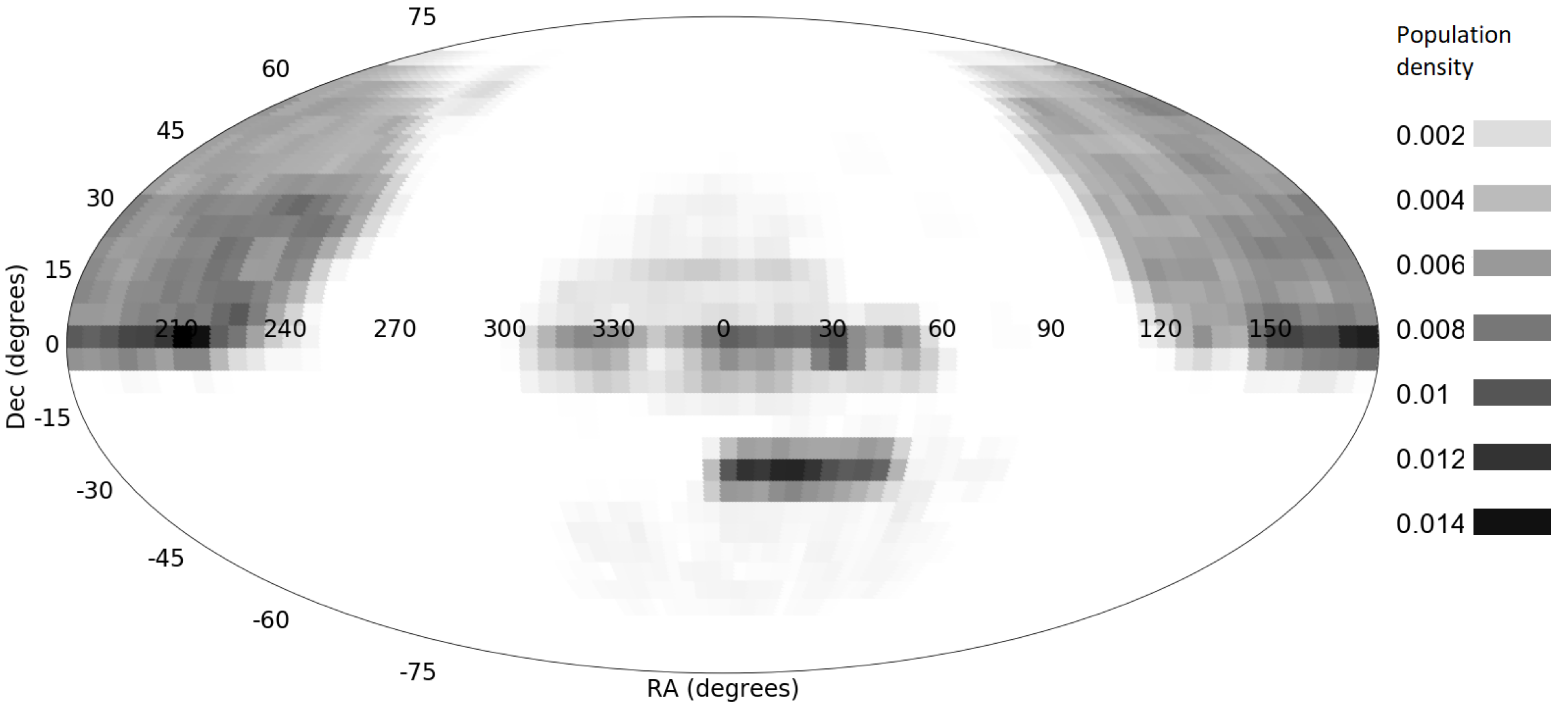}
\caption{The density of the galaxy distribution in the dataset. The density in each $5^o\times5^o$ field of the sky is determined by the number of galaxies in the $5^o\times5^o$ field, divided by the total number of galaxies in the dataset.}
\label{population}
\end{figure}

Combining all three datasets and removing objects that appeared in more than one dataset provided a dataset of 90,023 galaxies with spectra. Figure~\ref{z_distribution} shows the distribution of the redshift in the dataset, and Figure~\ref{ra_distribution} shows the distribution of the galaxies by their RA. Figure~\ref{population} shows the galaxy population density in each $5^o\times5^o$ field of the sky. The density is determined by the number of galaxies in each $5^o\times5^o$ field divided by the total number of galaxies.

An important property of the dataset used in this study is that the data are retrieved and analyzed without any prior assumptions, and are not based on any existing catalog of galaxy morphology. Because the dataset is made of data from several telescopes covering both the Northern and Southern hemispheres, there is no existing catalog of galaxy morphology that includes all galaxies used in this study. More importantly, using an existing catalog of galaxy morphology might expose the analysis to bias of a catalog that was not prepared with the normalization of spin direction in mind. For instance, morphology catalogs that were prepared manually can be biased by the human perception \citep{land2008galaxy,hayes2017nature}. Due to the complex nature of the human perception, these subtle but consistent biases are very difficult to quantify and correct. 

In the past two decades, catalogs of galaxy morphology were prepared by using machine learning algorithms, and specifically deep neural networks such as \citep{gravet2015catalog,perez2015morphologies,goddard2020catalog,cheng2021galaxy}. While these catalogs are prepared by using a computer software, the  rules used to make the annotation are determined automatically by using training sets of galaxies that were annotated manually. Therefore, the machine learning systems can still capture the perceptional biases of the humans who annotated the data that was used to train the machine learning system. Also, the rules used by these machine learning systems are complex non-intuitive rules generated automatically from the data by which the machine learning system is trained. Due to their complexity, it is very difficult to verify theoretically or empirically that these rules are fully symmetric, and do not lead to certain biases.

It might be reasonable to assume that if such bias existed, it would have been expected to be consistent throughout the sky, and would not exhibit inverse asymmetry in opposite hemispheres. However, due to the complex nature of these algorithms it is very difficult to prove such claim. For instance, it has been shown that by selecting a different training set, a deep neural network produces a slightly but consistently different catalog \citep{dhar2022systematic}. Therefore, using catalogs that were prepared for other purposes, and do not necessarily ensure that the algorithm is fully symmetric in terms of the spin directions of the galaxies, can introduce an additional source of bias that depends on the way the catalog was prepared, and might be carried on to the rest of the analysis.

As discussed in Section~\ref{ganalyzer}, the algorithm used to determine the spin direction of the galaxies can also perform the broad morphological classification of elliptical and spiral galaxies \citep{shamir2011ganalyzer}, and identify just galaxies with identifiable spin direction. Elliptical galaxies or other galaxies that their spin direction cannot be determined are not included in the analysis, and the removal of the galaxies is done by the same model-driven symmetric algorithm described in Section~\ref{ganalyzer}. Further discussion on the use of catalogs of galaxy morphology can be found in Section~\ref{morphology_catalogs}.

\section{Results}
\label{results}

A very simple way of observing the distribution of galaxy spin directions in a certain field in the sky is by comparing the number of galaxies spinning clockwise to the number of galaxies spinning counterclockwise. The statistical significance of the difference can be determined by using binomial distribution, such that the probability of a galaxy to spin clockwise or counterclockwise is assumed to be 0.5. The asymmetry {\it A} can be defined by $A=\frac{cw-ccw}{cw+ccw}$, where {\it cw} is the number of galaxies spinning clockwise, and {\it ccw} is the number of galaxies spinning counterclockwise. 

A galaxy that seems to spin clockwise to an observer on Earth would seem to spin counterclockwise if the observer was in the opposite side of the galaxy. Therefore, the asymmetry of the distribution of galaxy spin directions in one hemisphere is expected to be inverse in the opposite hemisphere. Perhaps the most simple way of separating the sky into two hemispheres is such that one hemisphere is $(0^o<\alpha<180^o)$, and the other hemisphere is $(180^o<\alpha<360^o)$. That separation is clearly very simple, and used in this case for the sake of simplicty. Table~\ref{hemispheres} shows the number of clockwise and counterclockwise galaxies in each hemisphere, and the binomial probability to have such difference by chance.

\begin{table}
\centering
\begin{tabular}{lcccc}
\hline
Hemisphere       & \# cw    & \# ccw    & $\frac{cw-ccw}{cw+ccw}$  & P \\
                 & galaxies & galaxies  &                          &   \\
\hline
$(0^o<\alpha<180^o)$    &   23,070    & 23,606  &   -0.0115      &  0.007  \\   
$(180^o<\alpha<360^o)$  &   21,808    & 21,539  &   0.0062      &  0.09   \\
\hline
\end{tabular}
\caption{Distribution of clockwise and counterclockwise galaxies in opposite hemispheres. The P values are the binomial distribution probability to have such difference or stronger by chance when assuming mere chance 0.5 probability for a galaxy to spin clockwise or counterclockwise.}
\label{hemispheres}
\end{table}

As the table shows, the hemisphere $(0^o<\alpha<180^o)$ has a higher number of galaxies spinning clockwise, while the opposite hemisphere has a higher number of galaxies spinning counterclockwise. While the sign of the asymmetry is inverse in the opposite hemisphere, the asymmetry in the hemisphere $(180^o<\alpha<360^o)$ is not necessarily statistically significant. However, even if assuming no asymmetry in that hemisphere, after applying a Bonferroni correction the the Bonferroni-corrected P value of the asymmetry in $(0^o<\alpha<180^o)$ is $\sim$0.014, which is statistically significant.  When repeating the analysis by using the mirrored images the results inverse, as expected due to the symmetric nature of the Ganalyzer algorithm that was used to annotate the images.

The inverse asymmetry of galaxy spin direction between opposite hemispheres has been done in the past with several telescopes such as SDSS \citep{shamir2020patterns,shamir2020pasa}, DECam \citep{shamir2021large,shamir2022new}, and Pan-STARRS \citep{shamir2020patterns,shamir2021large,shamir2022new}. For instance, as shown in \citep{shamir2022new}, separating the entire dataset of galaxies imaged by DECam, mostly in the Southern hemisphere, into two hemispheres provided significant difference between the number of galaxies spinning in opposite directions in the two opposite hemispheres.

Clearly, the simple separation to two opposite hemispheres is a very simple analysis, and a more thorough analysis will be provided later in this section. One of the aspects that can be tested is the relationship between the magnitude of the difference and the redshift. Table~\ref{hemispheres_z} shows the same analysis shown in Table~\ref{hemispheres} for the hemisphere $(0^o<\alpha<180^o)$, but after separating the galaxies into galaxies with redshift of $z<0.15$ and galaxies with redshift of $z>0.15$. 

\begin{table}
\begin{tabular}{lcccc}
\hline
Redshift       & \# cw    & \# ccw    & $\frac{cw-ccw}{cw+ccw}$  & P \\
range          & galaxies & galaxies  &                          &   \\
\hline
$z<0.15$    &   15,767    & 15,877  &   -0.0034      &  0.26  \\   
$z>0.15$    &   7,303     & 7,729   &   -0.0283      &  0.0002   \\
\hline
\end{tabular}
\caption{Distribution of clockwise and counterclockwise galaxies in the hemisphere $(0^o<\alpha<180^o)$ in different redshift ranges.}
\label{hemispheres_z}
\end{table}

As the table shows, in the higher redshift ranges the asymmetry becomes significantly stronger than when limiting the redshift to lower redshift ranges. The observation is in agreement with previous results using smaller datasets of galaxies with spectra \citep{shamir2020patterns}. A more detailed analysis that shows the consistent increase of the difference and the statistical signal with the redshift is described in \citep{shamir2020patterns}.

\subsection{Analysis of a dipole axis in different redshift ranges}
\label{dipole}

The analysis by separating the sky into two simple hemispheres has the advantage of simplicity, but might not provide a full accurate analysis of the location of the most probable axis around which the galaxies are aligned. The separation into two simple hemispheres is arbitrary, and while it shows evidence of asymmetry by suing very simple statistics, it does not allow to profile or identify the location of the most likely axis. A more comprehensive analysis of the presence of a dipole axis can be done by fitting the cosine of the angular distances of the galaxies from each possible $(\alpha,\delta)$ combination to their spin directions.

That analysis can be done by assigning the galaxies with their spin direction $d$, which is 1 for galaxies spinning clockwise, and -1 for galaxies spinning counterclockwise. The cosines of the angular distances $\phi$ is then $\chi^2$ fitted into $d\cdot|\cos(\phi)|$. From each possible $(\alpha,\delta)$ integer combination, the angular distance $\phi_i$ between $(\alpha,\delta)$ and each galaxy {\it i} in the dataset is computed. The $\chi^2$ from each $(\alpha,\delta)$ is determined by Equation~\ref{chi2}
\begin{equation}
\chi^2_{\alpha,\delta}=\Sigma_i \frac{(d_i \cdot | \cos(\phi_i)| - \cos(\phi_i))^2}{\cos(\phi_i)} ,
\label{chi2}
\end{equation}
where $d_i$ is the spin direction of galaxy {\it i}, and $\phi_i$ is the angular distance between galaxy {\it i} and $(\alpha,\delta)$.

For each $(\alpha,\delta)$ combination, the $\chi^2$ is computed with the real spin directions of the galaxies, and then computed 1000 times when $d_i$ is assigned with a random spin direction. Using the $\chi^2$ from 1000 runs, the mean and standard deviation of the $\chi^2$ when the spin directions are random is computed. Then, the $\sigma$ difference between the $\chi^2$ computed with the real spin directions and the mean $\chi^2$ computed with the random spin directions is used to determine the $\sigma$ of the $\chi^2$ fitness to occur by chance. That is done for each $(\alpha,\delta)$ integer combination in the sky to determine the probability of each $(\alpha,\delta)$ to be the center of the dipole axis \citep{shamir2012handedness,shamir2019large,shamir2020pasa,shamir2020patterns,shamir2021particles,shamir2021large,shamir2022new}. 

\begin{figure}
\centering
\includegraphics[scale=0.15]{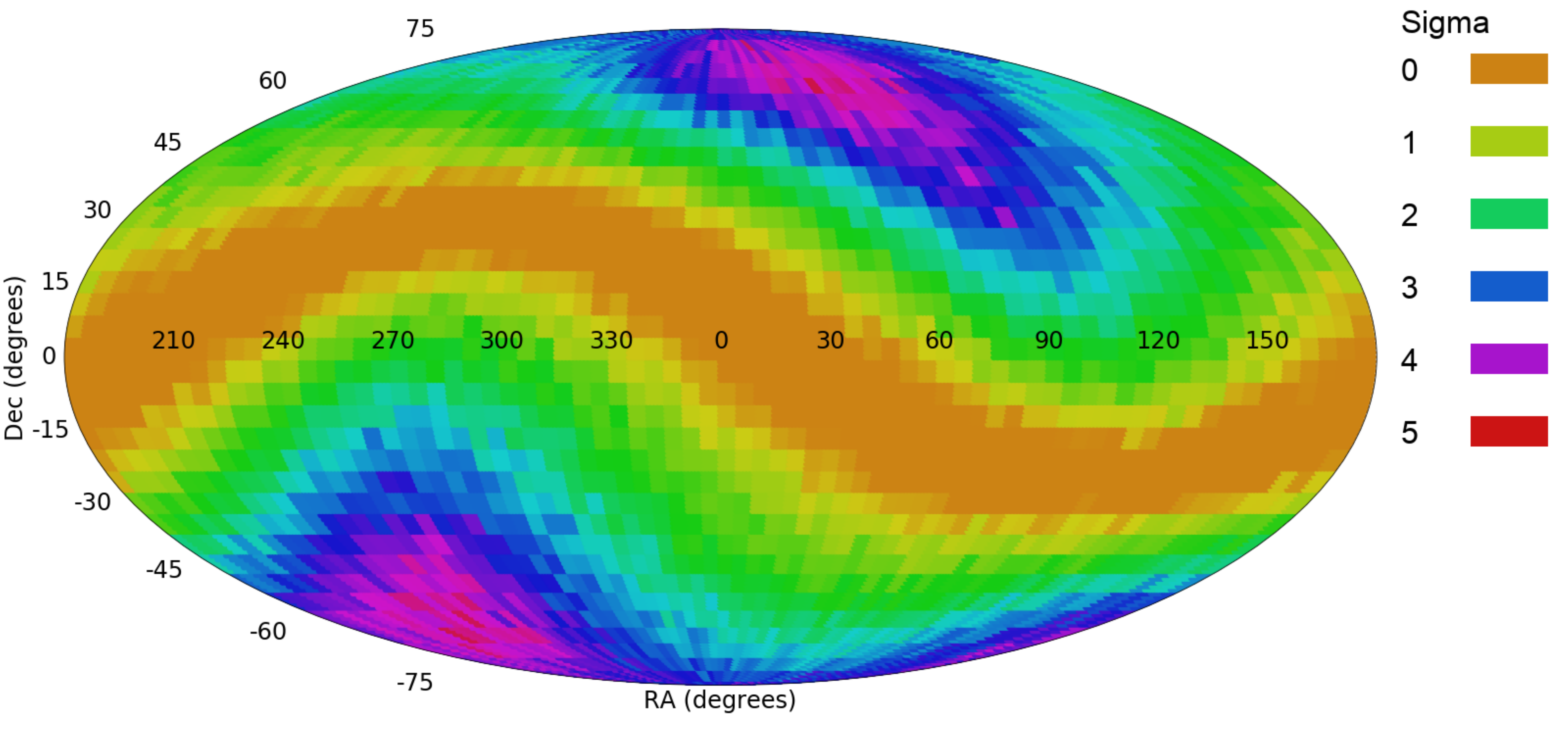}
\caption{The probability of a dipole axis in different $(\alpha,\delta)$ combinations.}
\label{dipole_all}
\end{figure}

Figure~\ref{dipole_all} shows the computed probabilities of a dipole axis in different $(\alpha,\delta)$ combinations. The most probable axis is at $(\alpha=65^o,\delta=52^o)$, with probability of 4.7$\sigma$ to occur by chance. The 1$\sigma$ error of the location of that axis is $(0^o,122^o)$ for the RA, and $(42^o,-77^o)$ for the declination. 

\begin{figure}[h]
\centering
\includegraphics[scale=0.15]{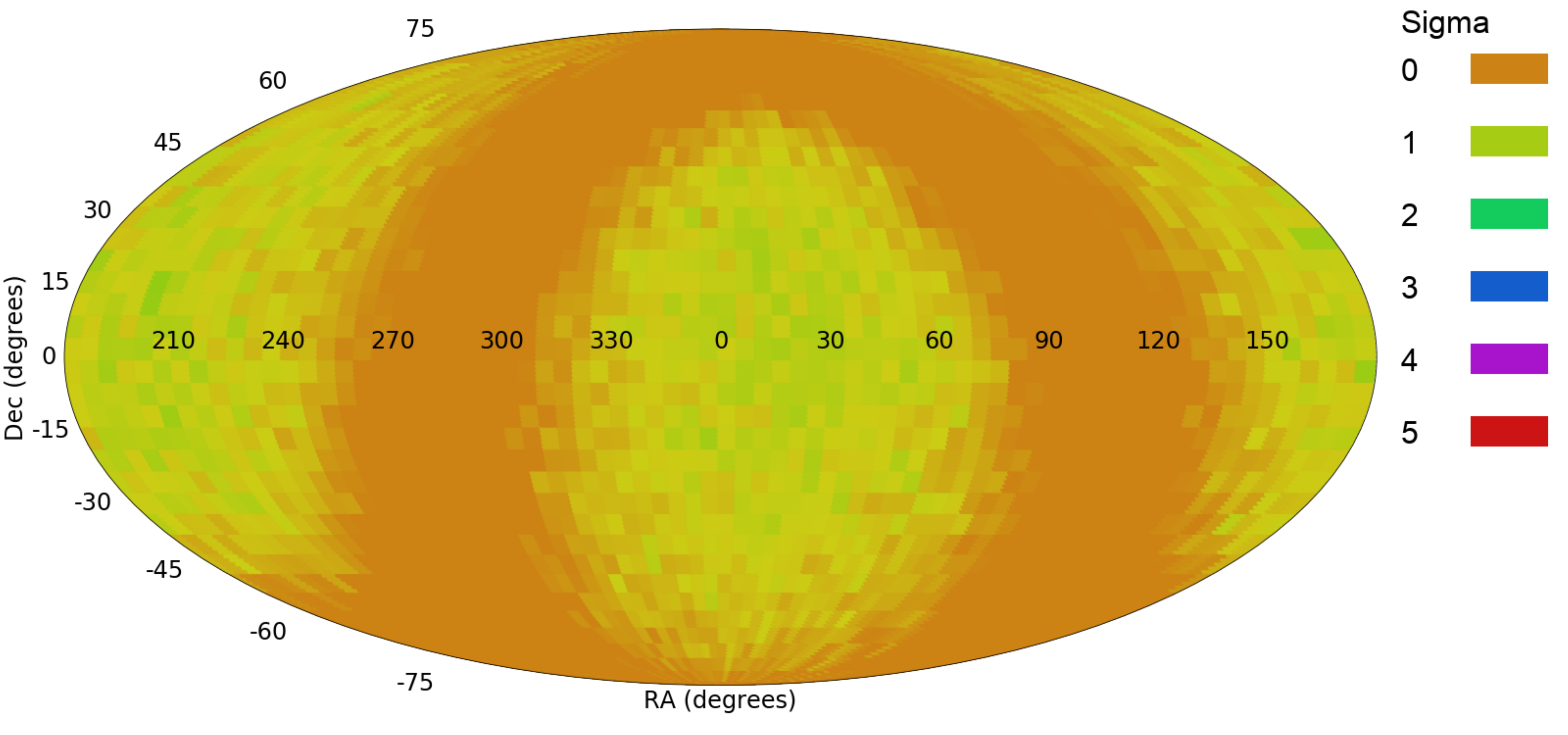}
\caption{The probability of a dipole axis at different $(\alpha,\delta)$ combinations when the galaxies are assigned with random spin directions.}
\label{z_all_random}
\end{figure}

When assigning the galaxies with random spin directions, the asymmetry becomes insignificant \citep{shamir2012handedness,shamir2019large,shamir2020pasa,shamir2020patterns,shamir2020large,shamir2021particles,shamir2021large,shamir2022new}. Figure~\ref{z_all_random} shows the probabilities of a dipole axis in different $(\alpha,\delta)$ combinations when using the same galaxies used in Figure~\ref{dipole_all}, but when assigning these galaxies with random spin directions. The most likely axis has statistical strength of 0.91$\sigma$.

\begin{figure*}[h]
\centering
\includegraphics[scale=0.16]{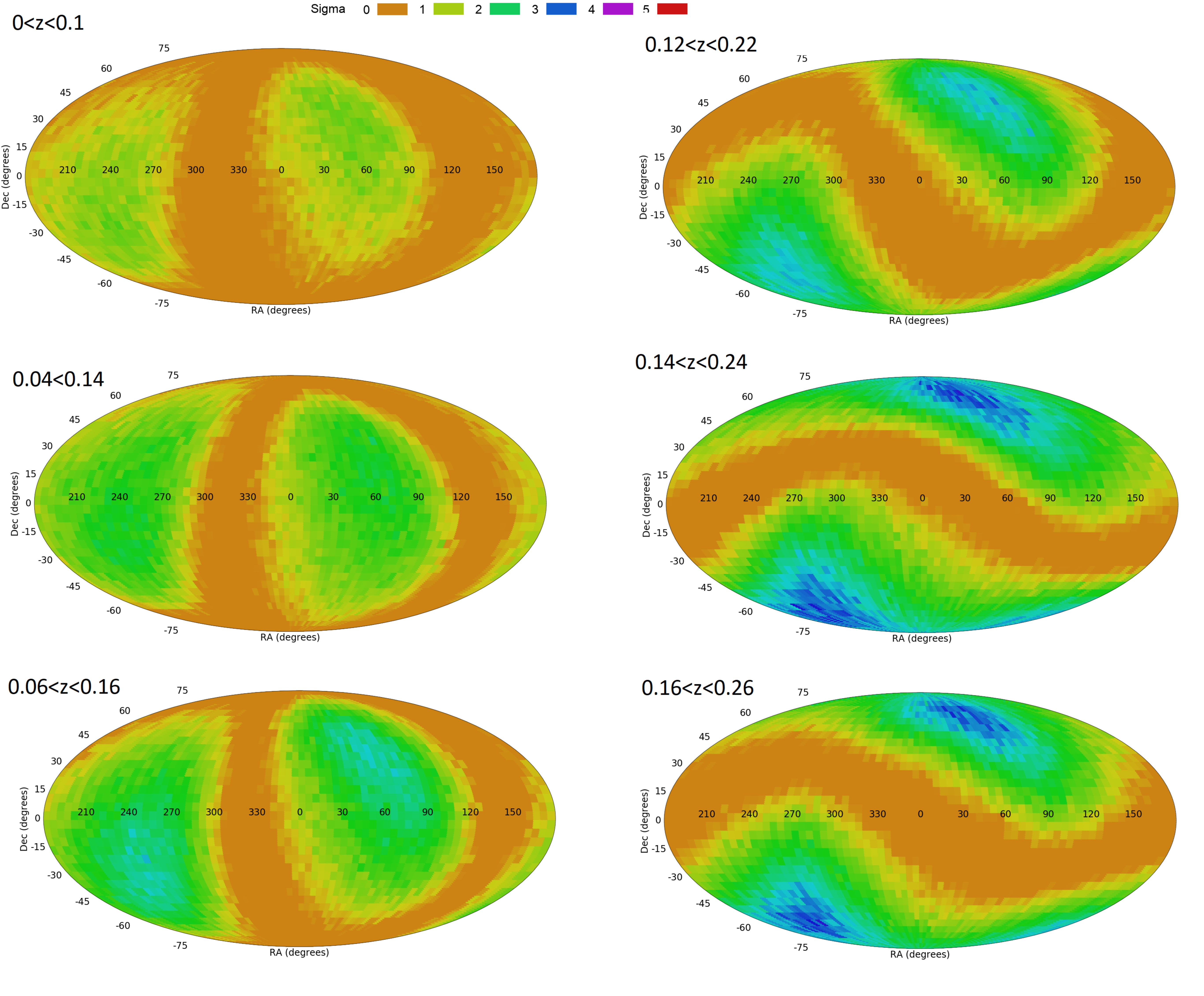}
\caption{The dipole axes in galaxy spin direction from different $(\alpha,\delta)$ combinations in different redshift ranges.}
\label{dipole_z}
\end{figure*}

Figure~\ref{dipole_z} shows the profile when fitting the galaxy spin directions into a dipole axis alignment using galaxies at different redshift ranges. The figure shows that the location of the most likely axis changes with the redshift, until around the redshift range of $0.12<z<0.22$, after which it stays constant. One immediate explanation to the change in the position of the most likely location of the axis when the redshift increases is that the axis does not necessarily go through Earth.

\begin{figure}[h]
\centering
\includegraphics[scale=0.25]{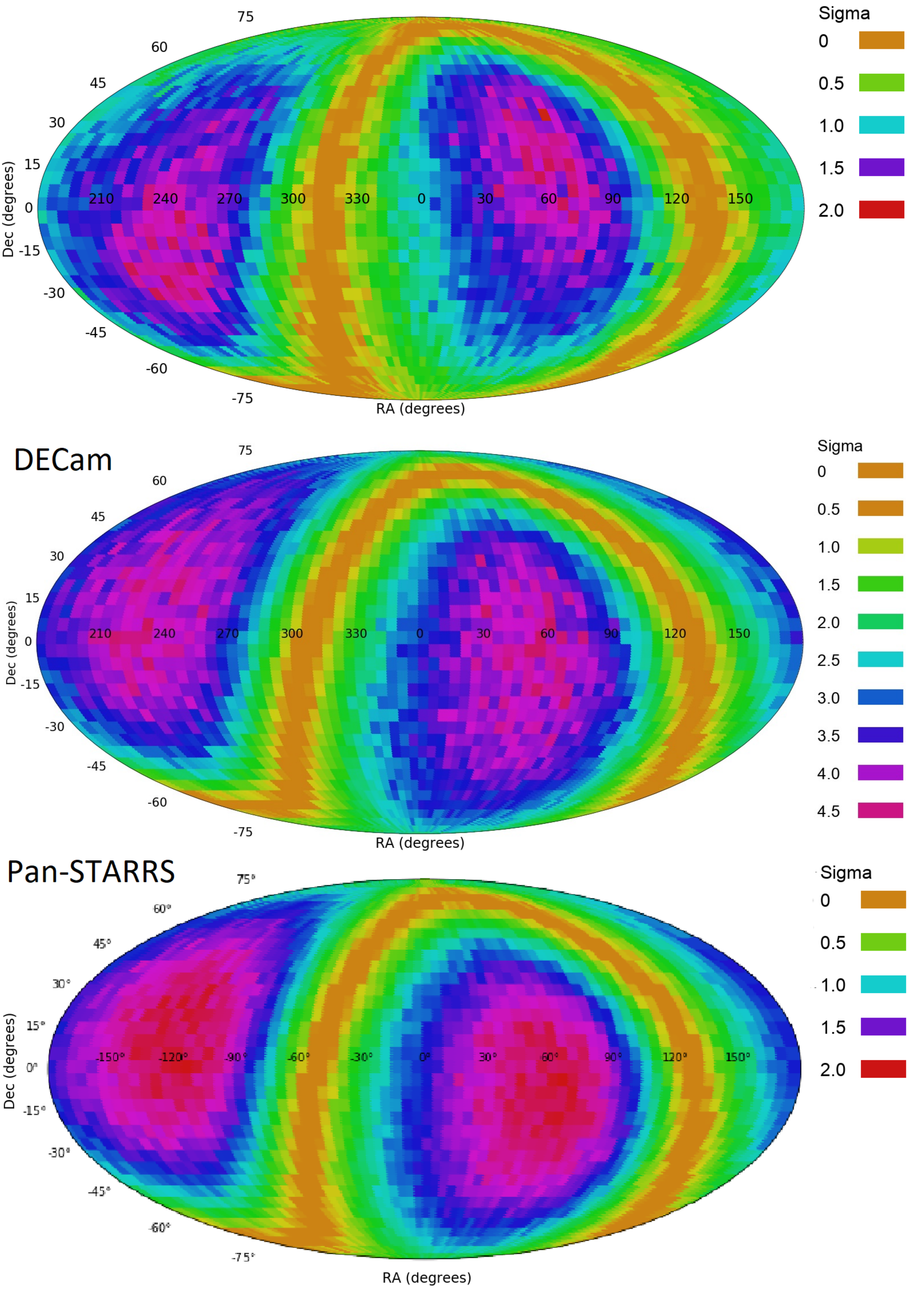}
\caption{The profile of galaxy spin direction distribution in the dataset described in Section~\ref{data} (top), compared to the profiles observed in Pan-STARRS (bottom) and DECam (middle) as reported in \citep{shamir2021large}. The graphs shows similar profiles of galaxy spin direction distribution observed in the different telescopes.}
\label{panstarrs_decam}
\end{figure}

The profiles of asymmetry observed with the low redshift galaxies can be compared to previous analyses of galaxy spin directions using Pan-STARRS \citep{shamir2020patterns} and DECam \citep{shamir2021large}. The Pan-STARRS dataset used in \cite{shamir2020patterns} includes 33,028 galaxies, and the DECam dataset contains 807,898 galaxies \citep{shamir2021large}. The vast majority of the galaxies in these datasets do not have redshift, but because the magnitude of all galaxies is limited to 19.5, the redshift is also expected to be relatively low. Figure~\ref{panstarrs_decam} shows the profile of asymmetry observed with Pan-STARRS and DECam as reported in \citep{shamir2021large,shamir2022new}. These results are compared to the results with the dataset described in Section~\ref{data} when 43,566 galaxies are selected such that the redshift distribution of these galaxies is similar to the redshift distribution of the DECam galaxies as determined by the redshift distribution of the few DECam galaxies with spectra. The agreement between different telescopes indicates that  the distribution of galaxy spin direction reflects the actual Universe rather than a certain unknown anomaly in a telescope.  

\begin{figure}[h]
\centering
\includegraphics[scale=0.15]{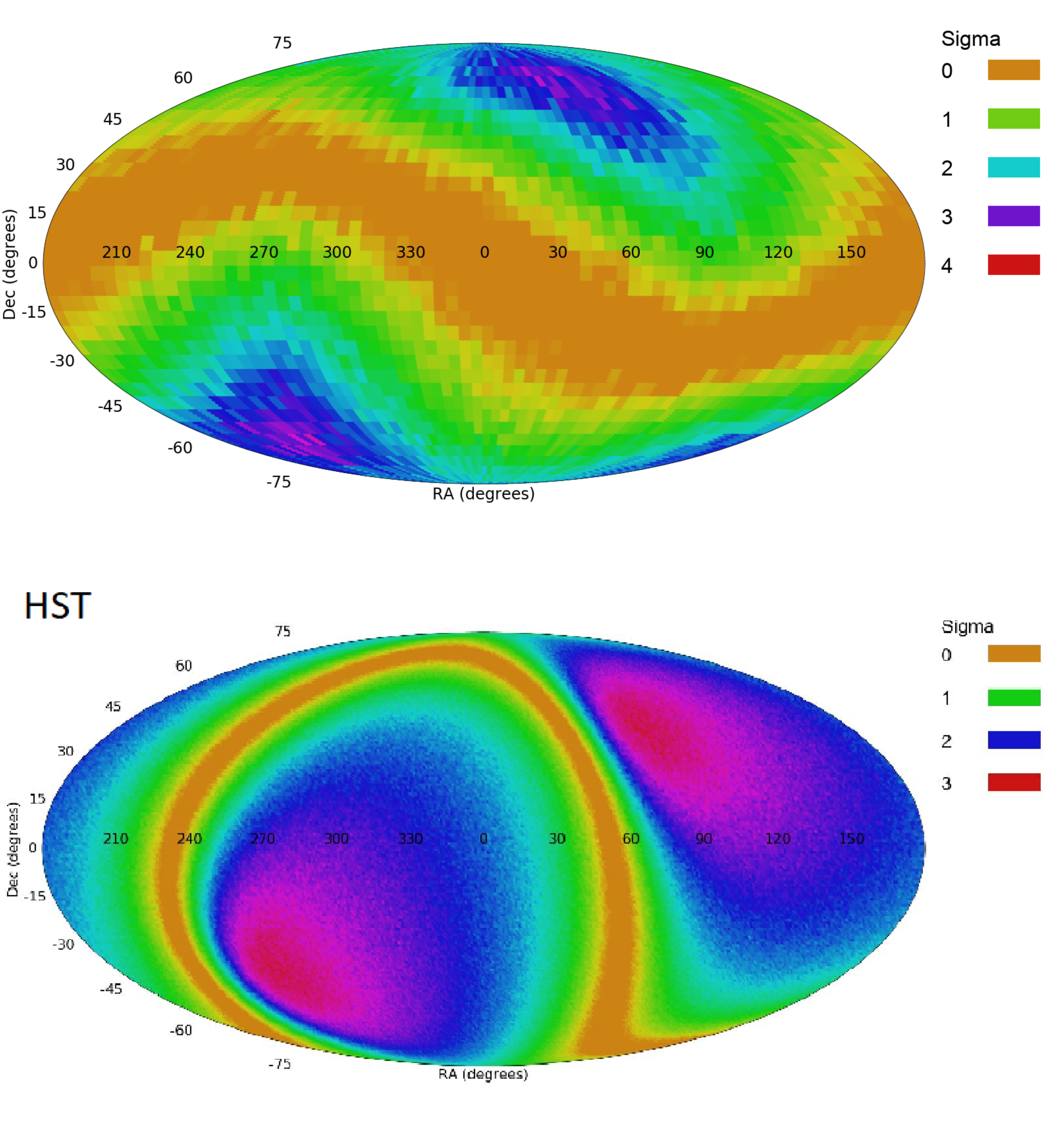}
\caption{The probability of a dipole axis in galaxy spin directions from different integer $(\alpha,\delta)$ combinations in HST (bottom) and the galaxies in the dataset described in Section~\ref{data} (top) when the redshift of the galaxies is limited to $0.16<z<0.26$.}
\label{sdss_hst}
\end{figure}

The asymmetry in the distribution of galaxy spin directions observed when the redshift of the galaxies is relatively high can be compared to the asymmetry profile observed by galaxies imaged by the Cosmic Assembly Near-infrared Deep Extragalactic Legacy Survey \citep{grogin2011candels,koekemoer2011candels} of the Hubble Space Telescope (HST). The HST dataset contains 8,690 galaxies annotated manually by their spin direction, as explained in \citep{shamir2020pasa}. During that process of annotation, a random half of the galaxies were mirrored for the first cycle of annotation, and then all galaxies were mirrored for a second cycle of annotation as described in \citep{shamir2020pasa} to offset possible effect of human perceptional bias. That provided a complete dataset that is also not subjected to atmospheric effects. A full description of the dataset and the analysis of the distribution of galaxy spin direction is described in \citep{shamir2020pasa}. These galaxies do not have spectra, but due to the nature of the Hubble Space Telescope it is clear that the galaxies imaged by HST are of much higher redshift than the galaxies imaged by the Earth-based sky surveys. Figure~\ref{sdss_hst} shows the profile observed using the HST galaxies as described in \citep{shamir2020pasa}, and the profile of asymmetry when using the dataset described in Section~\ref{data} when the redshift of the galaxies is limited to $0.16<z<0.26$. The most likely axis observed with HST galaxies is at $(\alpha=78^o,\delta=47^o)$, with probability of 2.8$\sigma$ to have such distribution by chance. As the figure shows, although the two datasets do not share any galaxies, they show similar positions of the most likely axes observed in the distribution of galaxy spin directions. The most likely axis observed with the dataset described in Section~\ref{data} peaks at $(\alpha=48^o,\delta=67^o)$.

Table~\ref{dipole_redshift_ranges} shows the location of the most likely axis when the dataset described in Section~\ref{data} is separated into different redshift ranges. As the table shows, the declination of the galaxies changes consistently as the redshift gets higher, while the change in the right ascension is milder. 

\begin{table}
\scriptsize
\begin{tabular}{|c|c|c|c|c|}
\hline
        z      & \# galaxies  &  RA                & Dec         & Statistical   \\ 
               &              & (degrees)       & (degrees)      & significance  \\  
\hline
0.0-0.1        &   41,218    &   61       &   10        &   1.3$\sigma$          \\				
0.01-0.11      &   45,618    &   56       &   22        &   1.2$\sigma$         \\				
0.02-0.12      &   49,693    &   53       &   30        &   1.6$\sigma$          \\				
0.03-0.13      &   51,027    &   55       &   32        &   1.7$\sigma$           \\				
0.04-0.14      &   51,243    &   62       &   36        &   1.9$\sigma$           \\				
0.05-0.15      &   50,446    &   48       &   45        &   2.1$\sigma$           \\				
0.06-0.16      &   48,362    &   54       &   46        &   2.4$\sigma$           \\				
0.07-0.17      &   44,739    &   60       &   49        &   3.2$\sigma$             \\				
0.08-0.18      &  40,101     &   53       &   52        &   3.4$\sigma$             \\				
0.09-0.19      &   35,565    &   58       &   55        &   4.0$\sigma$         \\				
0.1-0.2        &   32,206    &   61       &   53        &   3.6$\sigma$             \\				
0.11-0.21      &   28,486    &   61       &   50        &   2.9$\sigma$             \\				
0.12-0.22      &   24,462    &   54       &   60        &   2.7$\sigma$              \\				
0.13-0.23      &   21,001    &   58       &   58        &   2.8$\sigma$             \\				
0.14-0.24      &   17,497    &   55       &   65        &   3.2$\sigma$             \\	
0.15-0.25      &   14,785    &   57       &   64        &   3.3$\sigma$            \\	
0.16-0.26      &  12,664     &   48       &   67        &   3.1$\sigma$      \\				
\hline
\end{tabular}
\caption{The most likely axis when the galaxies are limited to different redshift ranges.}
\label{dipole_redshift_ranges}
\end{table}

\begin{figure}[h]
\centering
\includegraphics[scale=0.25]{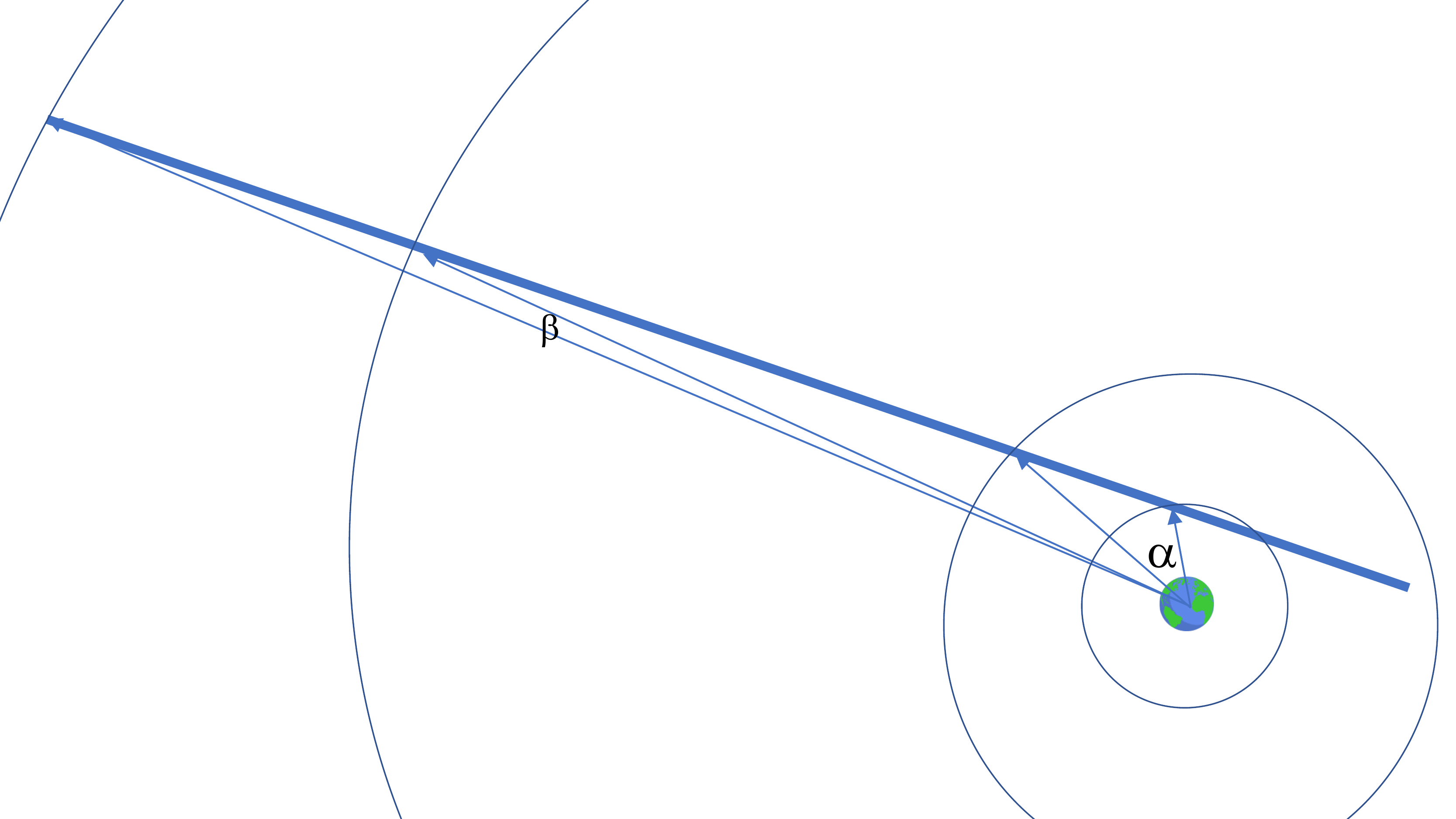}
\caption{Changes in the position of a non-Earth centered axis as seen from Earth when its peaks are identified at different redshifts. The change in the location of the axis in lower redshift range $\alpha$ is far larger than the change in the location in higher redshift range $\beta$.}
\label{restframe_axis}
\end{figure}

The change in the location of the peak of the most likely axis when the redshift of the galaxies changes can be viewed as an indication that the axis does not necessarily go directly through Earth. In such case, the location of the most likely axis is expected to change at low redshifts, and then to remain nearly constant at the higher redshifts. Figure~\ref{restframe_axis} displays a simple two-dimensional illustration of a possible axis compared to Earth. The angle $\alpha$ is measured between two points determined using two different redshifts, but the two redshifts are relatively low. That angle is much larger than $\beta$, which is the difference in the position of the dipole axis as seen from Earth when using two higher redshifts. Therefore, an axis that does not go directly through Earth is expected to change its location as seen from Earth in lower redshifts, but remain nearly at the same location when observed in higher redshift ranges.

\begin{figure}[h]
\centering
\includegraphics[scale=0.35]{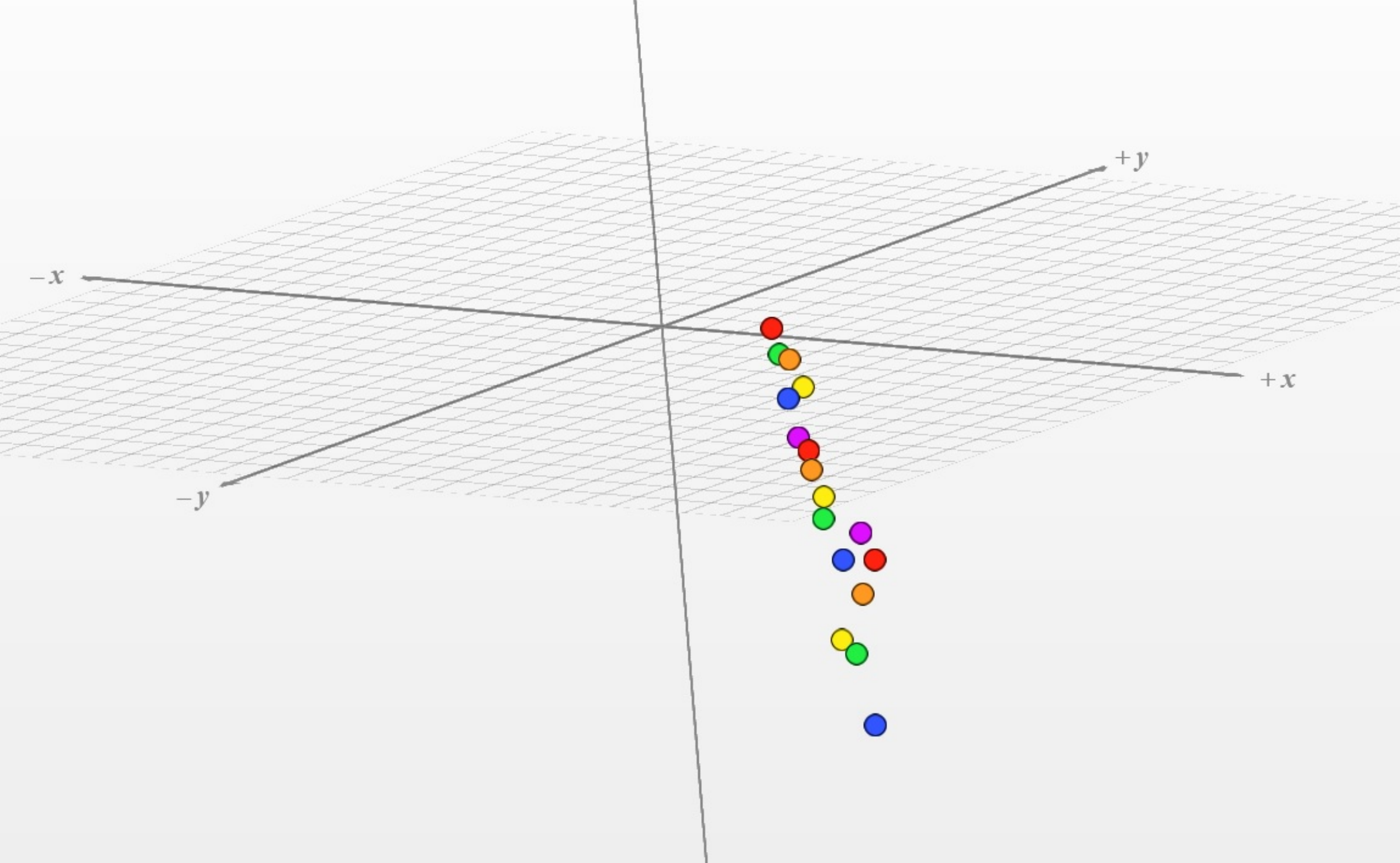}
\caption{Three-dimensional visualization of the points in Table~\ref{dipole_redshift_ranges}. The points form an axis that might not go directly through Earth.}
\label{dipole_axis_points}
\end{figure}

Figure~\ref{dipole_axis_points} visualizes the most likely axis points in a 3D space such that the distance {\it d} is determined by converting the mean redshift of the galaxies in each redshift range to the distance, measured in Mpc. The 3D transformation is then performed by Equation~\ref{transformation}. As expected, the points are aligned in a manner that forms a three-dimensional axis.

\begin{equation}
\begin{split}
x= & \cos(\alpha)\cdot d \cdot \cos(\delta)  \\
y= & \sin(\alpha)\cdot d \cdot \cos(\delta)  \\ 
z= & d \cdot sin(\delta)  \\
\label{transformation}
\end{split}
\end{equation}

To estimate the three dimensional direction of the axis, a simple analysis between two points $\eta$ and $\xi$ in a three-dimensional space can be used as shown by Equation~\ref{az_alt_from_3D}. Since the points in Table~\ref{dipole_redshift_ranges} as visualized in Figure~\ref{restframe_axis} are aligned in an axis, the direction of the axis can be determined by the two most distant points from each other. These two points are also the two points with the lowest and highest redshift as shown in Table~\ref{dipole_redshift_ranges}.

The closet point to Earth shown in Table~\ref{dipole_redshift_ranges} is at $(\alpha=61^o,\delta=10^o)$, and $\sim$283 Mpc away, based on the average 0.064 mean redshift of the galaxies in that redshift range. An observer in that point would see an axis in $(\alpha=262^o,\delta=85^o)$, and the other end of that axis in $(\alpha=82^o,\delta=-85^o)$.

\begin{equation}
\begin{split}
\alpha = & atan2(y_{\eta}-y_{\xi}, \sqrt{(x_{\eta}-x_{\xi})^2 + (z_{\eta}-z_{\xi})^2})  \\
\delta = & atan2(-(x_{\eta}-x_{\xi}),-(z_{\eta}-z_{\xi}))   \\
\label{az_alt_from_3D} 
\end{split}
\end{equation}

\subsection{Quadrupole axis analysis}
\label{quadrupole}

Analysis of the distribution of the CMB provided consistent evidence of a quadrupole axis alignment of the CMB anisotropy \citep{efstathiou2003low,feng2003double,cline2003does,weeks2004well,gordon2004low,piao2004suppressing,piao2005possible,campanelli2006ellipsoidal,campanelli2007cosmic,gruppuso2007complete,jimenez2007cosmology,rodrigues2008anisotropic,zhe2015quadrupole,santos2015influence}. Fitting the galaxy spin directions into quandrupole axis alignment can be done as described in Section~\ref{dipole}, but instead of $\chi^2$ fitting $\cos(\phi)$ into $d\cdot|\cos(\phi)|$ as was done for identifying a dipole axis, the quadrupole alignment is identified by $\chi^2$ fitting $\cos(2\phi)$ into $d\cdot|\cos(2\phi)|$, as described in \citep{shamir2019large,shamir2020patterns,shamir2021large}.

\begin{figure}
\centering
\includegraphics[scale=0.15]{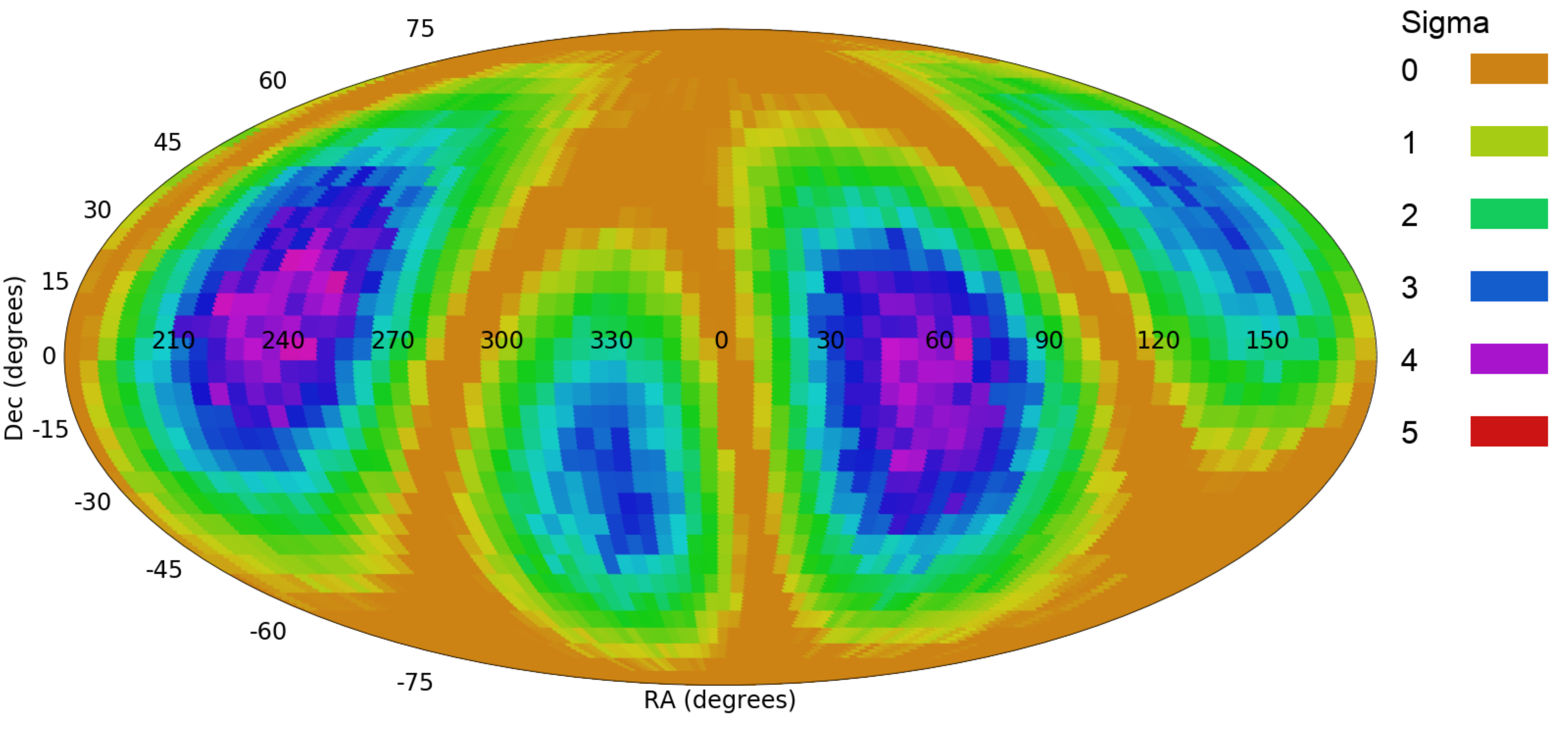}
\caption{The probability of a quadrupole axis in galaxy spin directions in different $(\alpha,\delta)$ combinations.}
\label{quad_all}
\end{figure}

\begin{figure}
\centering
\includegraphics[scale=0.15]{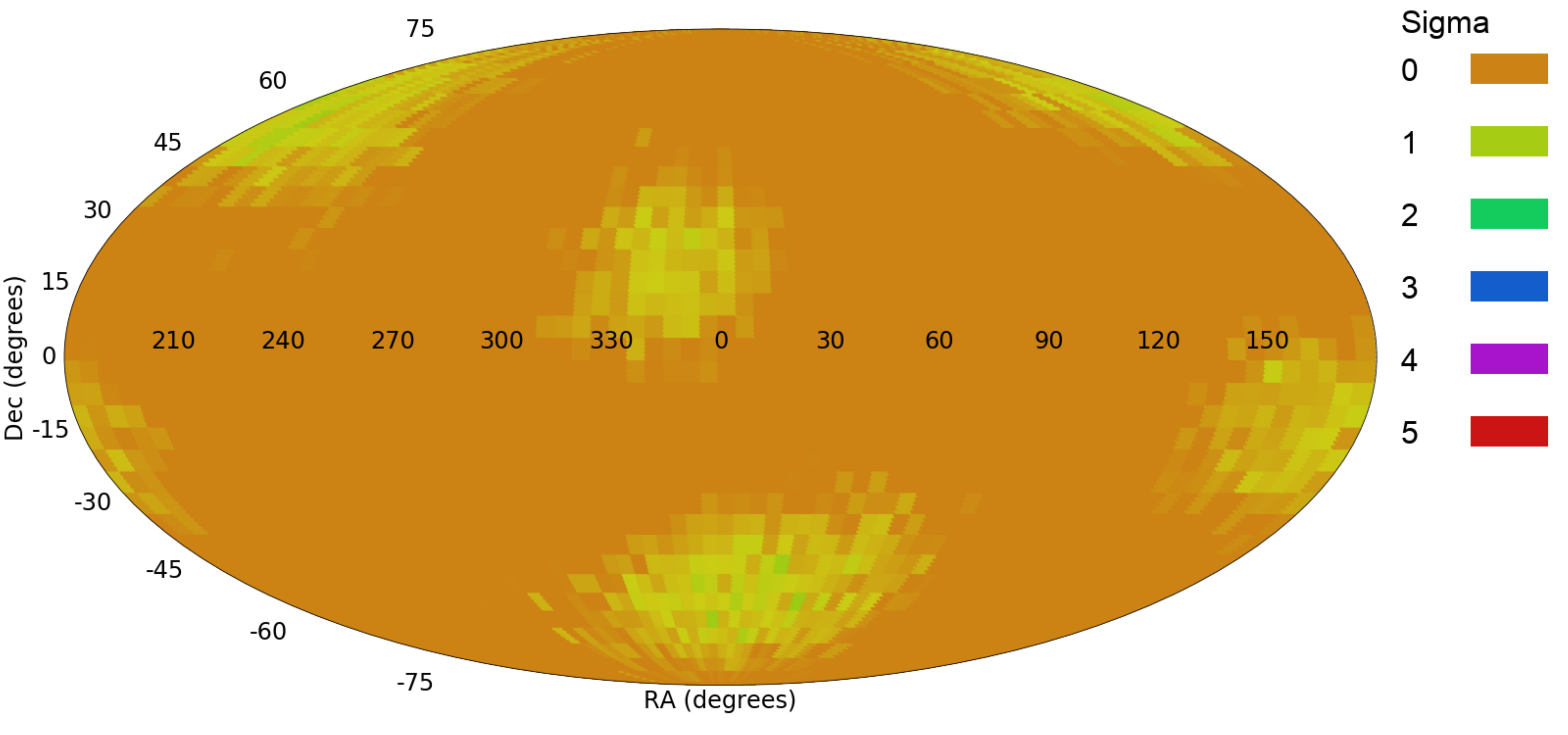}
\caption{The probability of a quadrupole axis in galaxy spin directions in different $(\alpha,\delta)$ combinations when the galaxies are assigned with random spin directions.}
\label{quad_random}
\end{figure}

Figure~\ref{quad_all} shows the likelihood of a quadrupole axis in different $(\alpha,\delta)$ combinations. One axis peaks at $(\alpha=52^o,\delta-8^o)$, with statistical signal of 4.3$\sigma$. The 1$\sigma$ error range for that axis is $(34^o,87^o)$ for the RA, and $(-42^o,36^o)$ for the declination. The other axis peaks at $(\alpha=151^o,\delta=31^o)$ with statistical signal of 3.1$\sigma$. The 1$\sigma$ error of that axis is $(118^o,191^o)$ for the RA, and $(-5^o,66^o)$ for the declination. Figure~\ref{quad_random} shows the same analysis when the galaxies are assigned random spin directions, showing that the axes disappear when the galaxy spin directions are random.

\begin{figure}[h]
\centering
\includegraphics[scale=0.25]{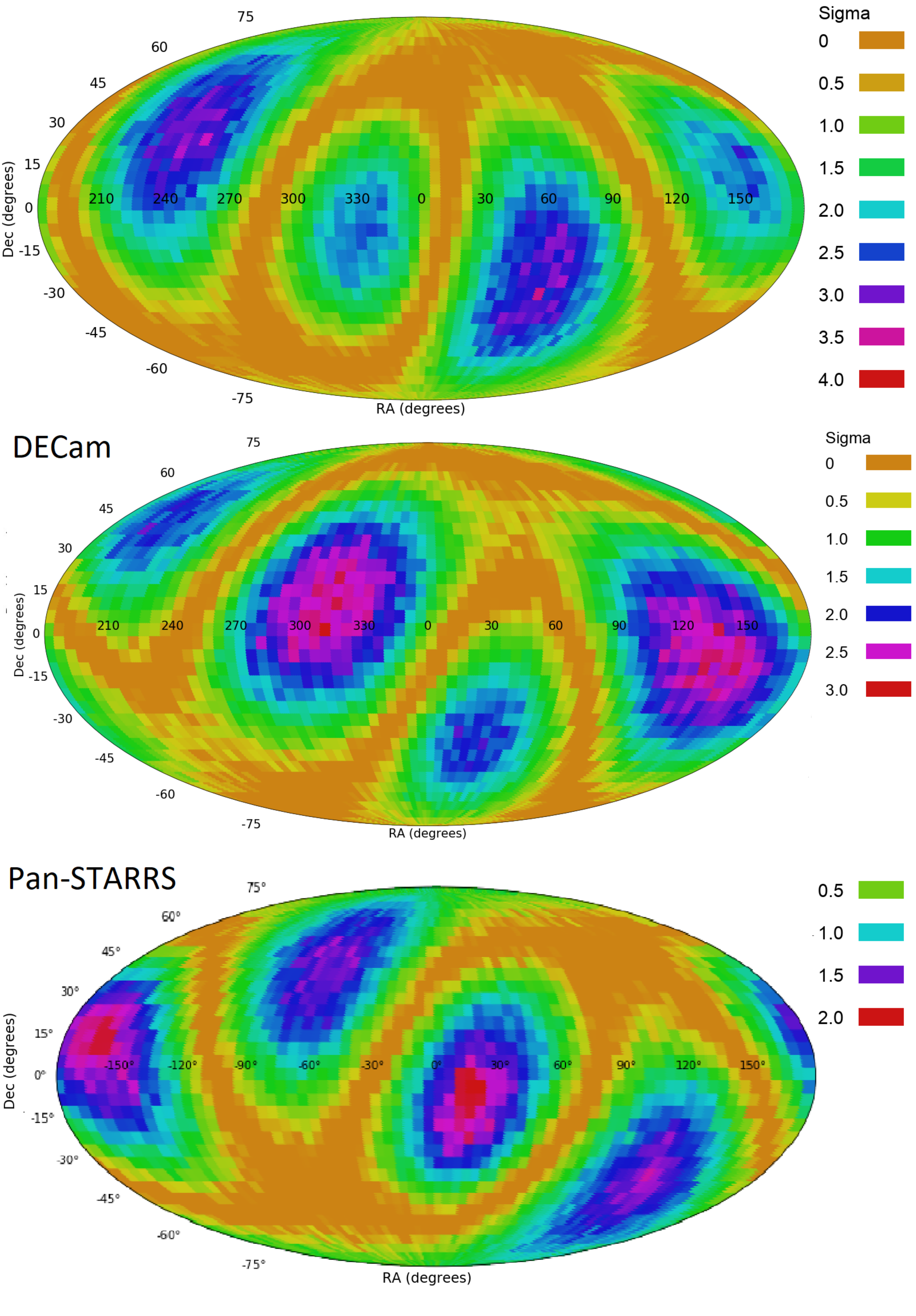}
\caption{The quadrupole profile of galaxy spin direction distribution in the dataset described in Section~\ref{data} (top), compared to the quadrupole profiles observed in Pan-STARRS (bottom) and DECam (middle) as reported in \citep{shamir2021large}.}
\label{panstarrs_decam_quad}
\end{figure}

As with the analysis of the dipole alignment, the analysis shown in Figure~\ref{quad_all} can be compared to previous analyses \citep{shamir2020patterns,shamir2021large}. Figure~\ref{panstarrs_decam_quad} displays the same analysis with Pan-STARRS and DECam as shown in \citep{shamir2021large}, compared to the analysis of the dataset described in Section~\ref{data}, using the 43,566 that have the same redshift distribution as the subset of DECam galaxies with redshift. The figure shows that the profile of galaxy spin direction distribution observed with the dataset described in Section~\ref{data} is similar to the previous results using Pan-STARRS and DECam shown in \citep{shamir2021large}.

\begin{figure*}
\centering
\includegraphics[scale=0.15]{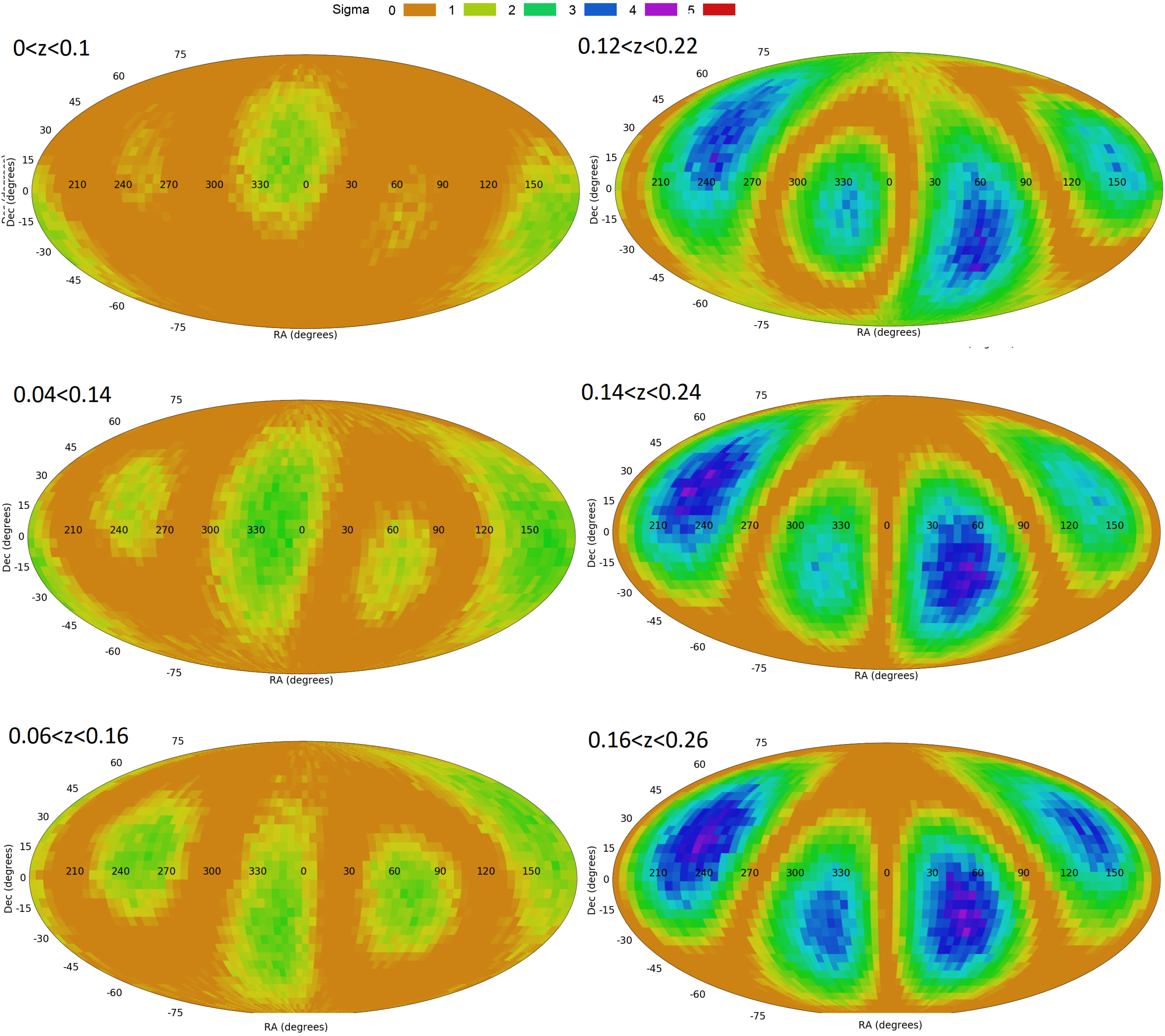}
\caption{The probability of a quadrupole axes at different $(\alpha,\delta)$ combinations in different redshift ranges.}
\label{quad_z}
\end{figure*}

Figure~\ref{quad_z} shows the analysis when fitting the distribution of galaxy spin direction to quadrupole axis alignment as observed when separating the redshift of the galaxies to certain different redshift ranges. As with the dipole alignment, the statistical signal of the quadrupole alignment becomes stronger as the redshift gets higher. The locations of the most likely axes change at the relatively lower redshifts, indicating that the axis is not necessarily Earth-centered. Table~\ref{quad_redshift_ranges} shows the most likely position of the two axes in each redshift range.

\begin{table*}
\scriptsize
\begin{tabular}{|c|c|c|c|c|c|c|c|}
\hline
        z      & \# galaxies  &  RA       & Dec          & Statistical     & RA       & Dec       & Statistical \\ 
               &                   &  (Axis 1) & (Axis 1)    & significance  & (Axis 2) & (Axis 2) & significance \\  
\hline
0.0-0.1        &   41,218    &   347      &   26        &   1.2$\sigma$  &   63    & -12      & 0.8$\sigma$     \\   
0.01-0.11      &   45,618    &   344      &   23        &   1.2$\sigma$  &   61    &  -12        & 0.8$\sigma$     \\		
0.02-0.12      &   49,693    &   342      &   21        &   1.2$\sigma$  &   61    &  -14     & 0.9$\sigma$     \\   
0.03-0.13      &   51,027    &   345      &   20        &   1.3$\sigma$  &    65    &  -15        & 1.1$\sigma$     \\ 
0.04-0.14      &   51,243    &   341      &   18        &   1.6$\sigma$  &    63   &  -17     & 1.2$\sigma$     \\  			
0.05-0.15      &   50,446    &   348      &   15        &   1.4$\sigma$  &    68   &  -17     & 1.2$\sigma$     \\  			
0.06-0.16      &   48,362    &   348      &   -27       &   1.5$\sigma$  &    71   &  -20     & 1.5$\sigma$     \\  
0.07-0.17      &   44,739    &   350      &   -36       &   1.6$\sigma$  &   66   &   -25    & 1.5$\sigma$     \\  
0.08-0.18      &  40,101     &   353      &   -40       &   1.5$\sigma$  &   67      &   -33   & 1.7$\sigma$     \\  			
0.09-0.19      &   35,565    &   344       &   -38      &  1.8$\sigma$   &     70    &    -34      & 2.1$\sigma$     \\  
0.1-0.2        &   32,206    &   349       &   -35        &   2.2$\sigma$  &     64    &    -37      & 2.6$\sigma$     \\  
0.11-0.21      &   28,486    &   351       &   -37        &   2.7$\sigma$  &    64    &    -36      & 2.9$\sigma$     \\  			
0.12-0.22      &   24,462    &   347       &   -34        &   2.9$\sigma$  &    62    &    -34      & 3.4$\sigma$     \\  				
0.13-0.23      &   21,001    &   350       &   -36        &   3.2$\sigma$  &    60     &    -37      & 3.7$\sigma$     \\  				
0.14-0.24      &   17,497    &   345      &   -35        &   3.5$\sigma$  &     61    &    -35      & 3.9$\sigma$     \\  
0.15-0.25      &   14,785    &   341       &   -32        &   3.9$\sigma$  &    57     &    -37      & 4.1$\sigma$     \\  
0.16-0.26      &  12,664     &   338       &   -36        &   3.1$\sigma$  &   65      &   -36       & 3.9$\sigma$     \\  			
\hline
\end{tabular}
\caption{The most likely quadrupole axis when the galaxies are limited to different redshift ranges.}
\label{quad_redshift_ranges}
\end{table*}

\begin{figure}[h]
\centering
\includegraphics[scale=0.45]{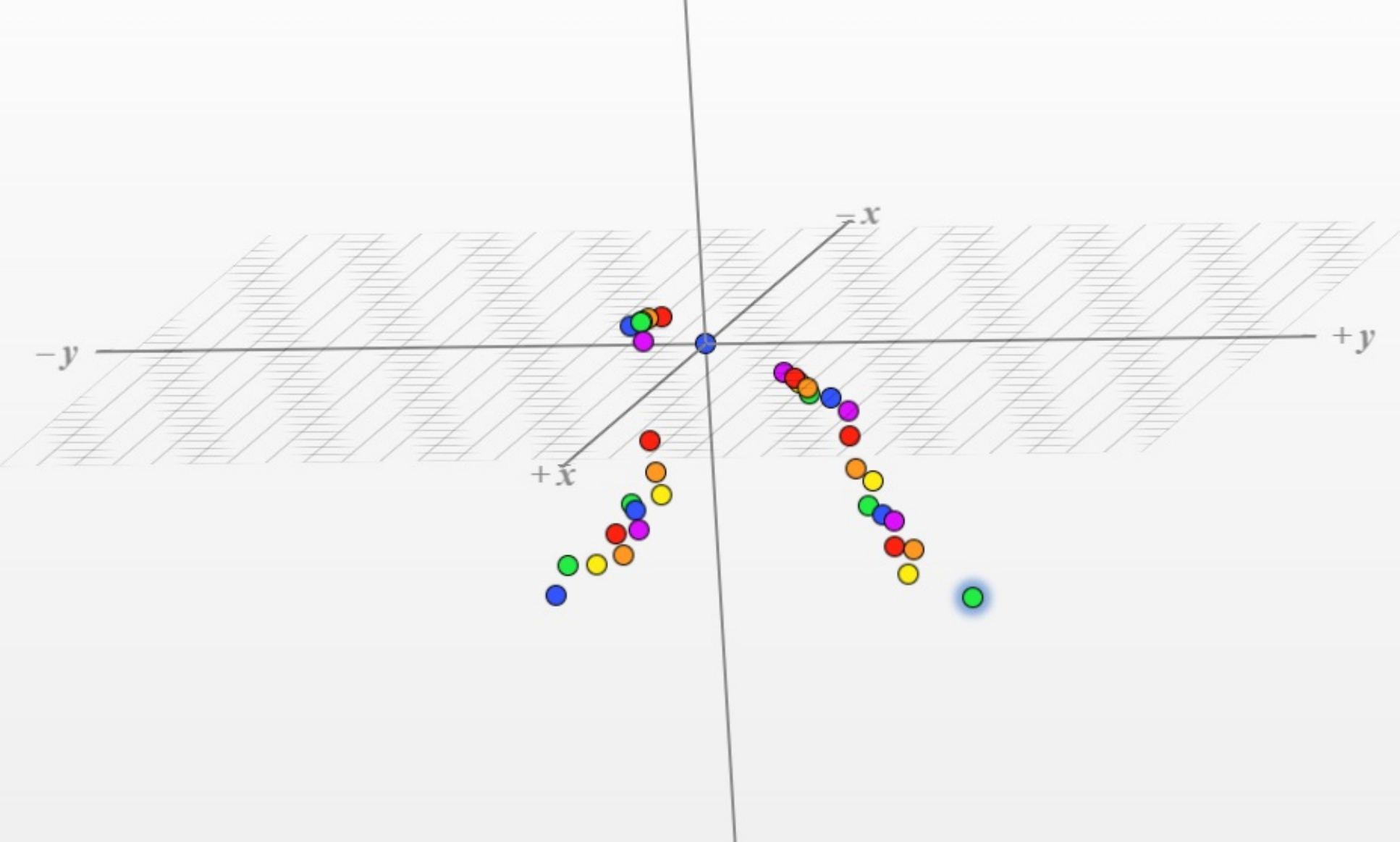}
\caption{Three dimensional visualization of the points in Table~\ref{quad_redshift_ranges}.}
\label{quad_axis_points}
\end{figure}
\begin{figure}[h]
\centering
\includegraphics[scale=0.45]{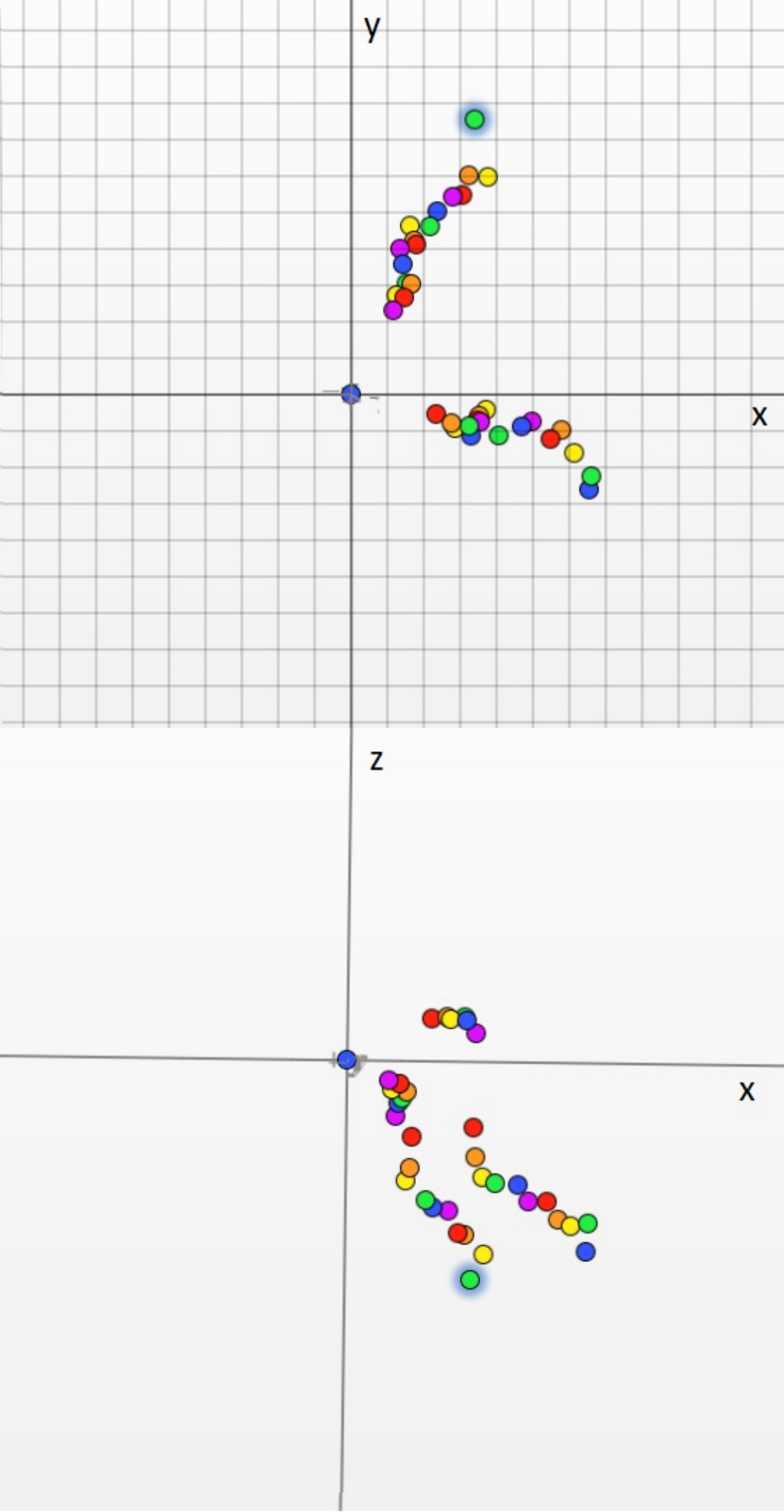}
\caption{The X,Y two-dimensional projection of the points in Table~\ref{quad_redshift_ranges} (top), and the X,Z two-dimensional projections.}
\label{x_y_z_points}
\end{figure}

Figure~\ref{quad_axis_points} displays the points in Table~\ref{quad_redshift_ranges} in a three dimensional space. The figure shows that the two axes form two lines that can be considered straight lines. Figure~\ref{x_y_z_points} shows the two-dimensional projections of the x,y plane and the x,z plane. The directions of the two axes can be determined by applying a linear regression to the points of each axis. According to the alignment of the points in Table~\ref{quad_redshift_ranges}, the two axes meet at RA 313$^o$ and declination of 65$^o$ from Earth, and at distance of 1736 Mpc.

\subsection{Analysis of differences in galaxy morphology}
\label{morphology}

\begin{figure}[h]
\centering
\includegraphics[scale=0.4]{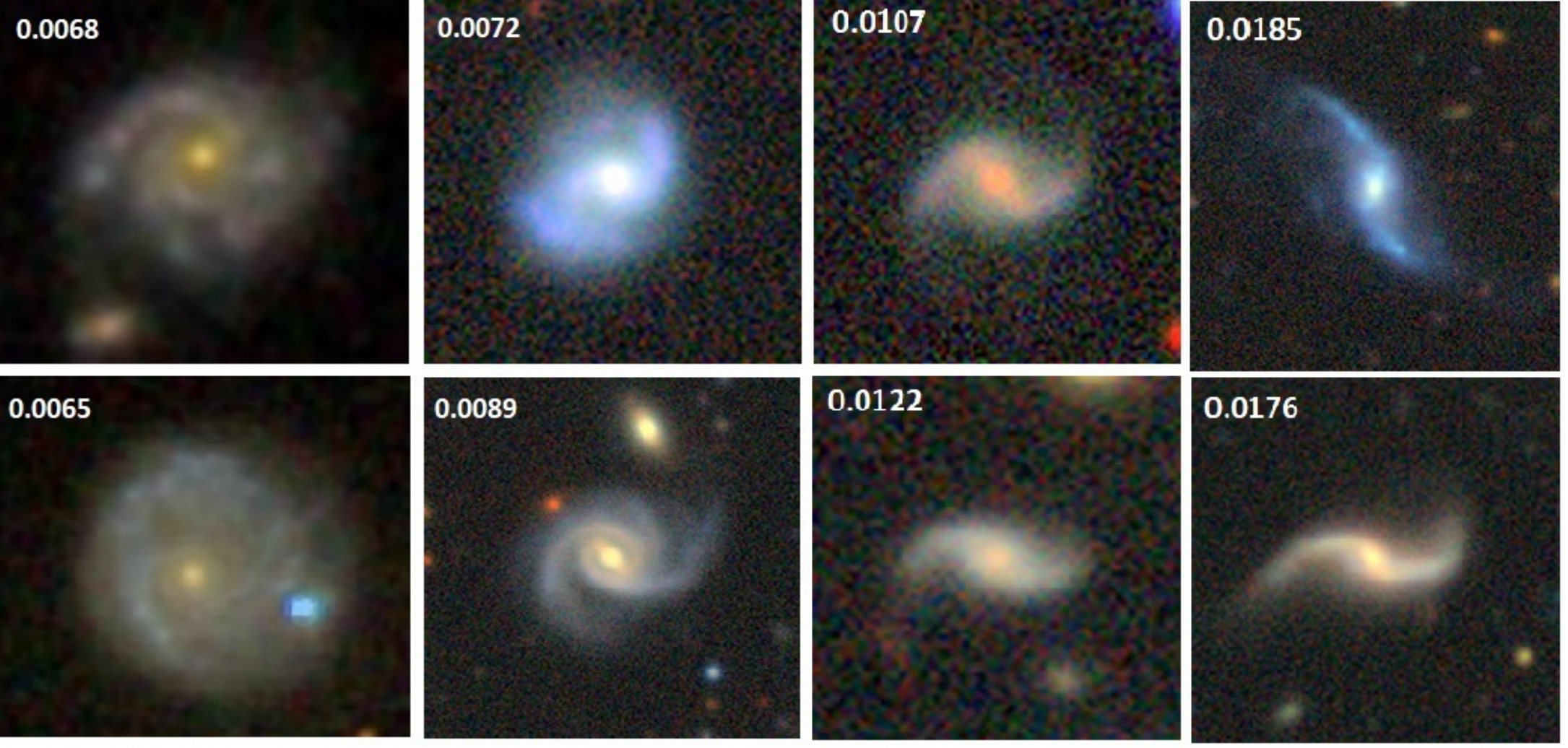}
\caption{$\frac{r}{S}$ of different galaxies. Higher $\frac{r}{S}$ indicates that the galaxy is less dense. The galaxies on the left have a relatively low $\frac{r}{S}$ of 0.0068 and 0.0065, indicating that the galaxies are more dense. The more sparse galaxies on the right have a higher $\frac{r}{S}$ of 0.0185 and 0.0176.}
\label{morphology_type}
\end{figure}

Another analysis was focused on differentiating the morphological differences between galaxies. For that purpose, the galaxies were separated by their $\frac{r}{S}$, where $r$ is the radius (in pixels), and $S$ is the galaxy surface size, also measured in pixels as described in \citep{shamir2011ganalyzer}. The surface size of the galaxy is measured by the number of foreground pixels, which can be counted after separating the foreground and background pixels of the galaxy as described in \citep{shamir2011ganalyzer}. That allowed to separate the galaxies by their mass distribution. A smaller $\frac{r}{S}$ value indicates that the surface size of the galaxy is large compared to its radius, and therefore the galaxy is more dense. Figure~\ref{morphology_type} shows examples of different galaxies and their $\frac{r}{S}$.

To test for the possible effect of galaxy morphology on the analysis, the data was separated into two subsets of similar size, such that in one subset all galaxies had $\frac{r}{S}\leq0.01$, and in the other subset all galaxies satisfied $\frac{r}{S}>0.01$. The subsets contained 45,526 and 44,497 galaxies, respectively. That led to two subsets that differ from each other by their mass distribution.

\begin{figure}
\centering
\includegraphics[scale=0.26]{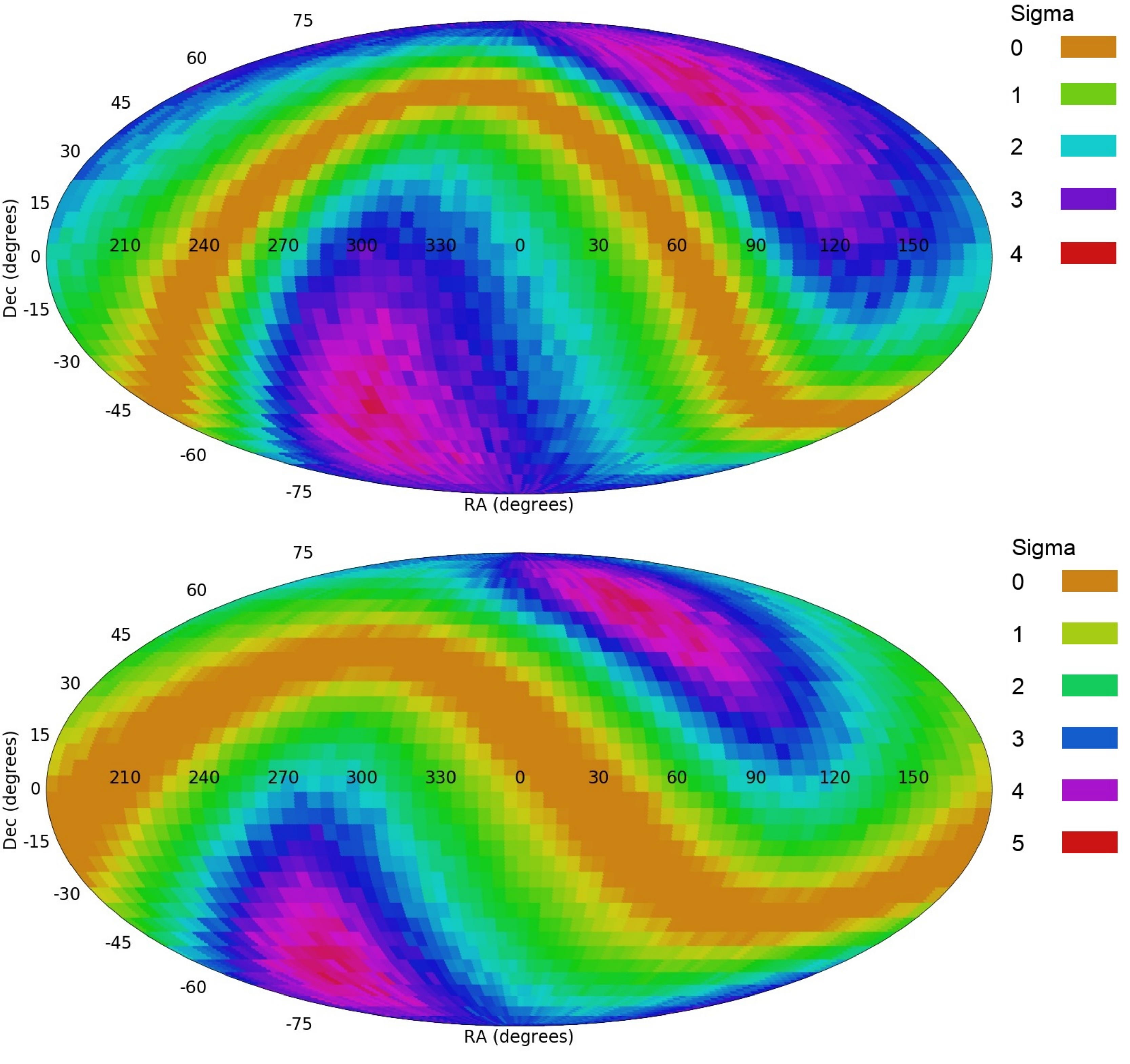}
\caption{The probability of a dipole axis in different $(\alpha,\delta)$ combinations when the galaxies have $\frac{r}{S}\leq0.01$ (top), and $\frac{r}{S}>0.01$ (bottom).}
\label{dipole_morphology}
\end{figure}

Figure~\ref{dipole_morphology} shows the results of the analysis described in Section~\ref{dipole}, such that the galaxies in the dataset are separated to galaxies with $\frac{r}{S}\leq0.01$ (top) and galaxies with $\frac{r}{S}>0.01$ (bottom). The figure shows that the two subsets of galaxies show a dipole axis at roughly the same location. When using galaxies that satisfy $\frac{r}{S}>0.01$, the statistical strength of the axis is 4.6$\sigma$. When the galaxies are limited to $\frac{r}{S}\leq0.01$, the statistical significance is 3.5$\sigma$. The 1.1$\sigma$ difference might provide certain indication of a link between the observed asymmetry and morphology of the galaxies. 

\section{Analysis of reasons that can lead to asymmetry not originated from the real sky}
\label{error}

While several messengers have shown evidence of violation of the cosmological-scale isotropy assumption as discussed in Section~\ref{introduction}, the null hypothesis would be that the spin directions of spiral galaxies are distributed randomly. This section discusses several possible reasons that could have led to the observation of asymmetry that does not reflect an asymmetry in the local Universe.



\subsection{Error in the galaxy annotation algorithm}

An error in the annotation algorithm can lead to any form of distribution, depends on the nature of the error. However, multiple indications show that the asymmetry cannot be the result of an error in the classification algorithm. The algorithm is a model-driven symmetric method that is based on clear rules. It is not based on complex data-driven rules used by pattern recognition systems. Such systems are complex and non-intuitive, rely on the data they are trained by and even by the order of the samples in the training set, making it virtually impossible to verify their symmetricity. The algorithm used here is fully symmetric, and follows clear defined rules as discussed in Section~\ref{data}.

Another evidence that the asymmetry is not driven by an error in the annotation algorithm is that the asymmetry changes between different parts of the sky, and inverse between opposite hemispheres. Since each galaxy is analyzed independently, a bias in the annotation algorithm is expected to be consistent throughout the sky, and it is not expected to flip in opposite hemispheres. The downloading of the images and the automatic analysis of the images were all done by the same computer, to avoid unknown differences between computers that can lead to bias or unknown differences in the way galaxy images are analyzed. 

Due to the theoretical and empirical evidence that the algorithm is symmetric, an error in the galaxy annotation is expected to impact clockwise and counterclockwise galaxies in a similar manner. If the galaxy annotation algorithm had a certain error in the annotation of the galaxies, the asymmetry {\it A} can be defined by Equation~\ref{asymmetry}.
\begin{equation}
A=\frac{(N_{cw}+E_{cw})-(N_{ccw}+E_{ccw})}{N_{cw}+E_{cw}+N_{ccw}+E_{ccw}},
\label{asymmetry}
\end{equation}
where $E_{cw}$ is the number of galaxies spinning clockwise incorrectly annotated as counterclockwise, and $E_{ccw}$ is the number of galaxies spinning counterclockwise incorrectly annotated as spinning clockwise. Because the algorithm is symmetric, the number of counterclockwise galaxies incorrectly annotated as clockwise is expected to be roughly the same as the number of clockwise galaxies missclassified as counterclockwise, and therefore $E_{cw} \simeq E_{ccw}$ \citep{shamir2021particles}. Therefore, the asymmetry {\it A} can be defined by Equation~\ref{asymmetry2}.

\begin{equation}
A=\frac{N_{cw}-N_{ccw}}{N_{cw}+E_{cw}+N_{ccw}+E_{ccw}}
\label{asymmetry2}
\end{equation}

Since $E_{cw}$ and $E_{ccw}$ cannot be negative, a higher rate of incorrectly annotated galaxies is expected to make {\it A} lower. Therefore, incorrect annotation of galaxies is not expected to lead to asymmetry, and can only make the asymmetry lower rather than higher.

An experiment \citep{shamir2021particles} of intentionally annotating some of the galaxies incorrectly showed that even when an error is added intentionally, the results do not change significantly even when as many as 25\% of the galaxies are assigned with incorrect spin directions, as long as the error is added to both clockwise and counterclockwise galaxies \citep{shamir2021particles}. But if the error is added in an asymmetric manner, even a small asymmetry of 2\% leads to a very strong asymmetry, and a dipole axis that peaks exactly at the celestial pole \citep{shamir2021particles}.

It should be mentioned that in one of the datasets shown here, which is the dataset acquired by HST, the annotation was done manually, and without using any automatic classification. The galaxies imaged by HST were annotated manually, and the results are in agreement with the automatic annotation of galaxies imaged by the other telescopes and annotated automatically.

\subsection{Bias in the sky survey hardware or photometric pipelines}

Autonomous digital sky surveys are some of the more complex research instruments, and involve sophisticated hardware and software to enable the collection, storage, analysis, and accessibility of the data. It is difficult to think of an error in the hardware or software that could lead to asymmetry between the number of galaxies spinning in opposite directions, but due to the complexity of these systems it is also difficult to prove that such error does not exist. That possible error is addressed here by comparing the results using data from several different telescopes. The instruments used in this study are independent from each other, and have different hardware and different photometric pipelines. As it is unlikely to have such bias in one instrument, it is very difficult to assume that all of these different instruments have such bias, and the profile created by that bias is consistent across all of them.

\subsection{Cosmic variance}

The distribution of galaxies in the universe is not completely uniform. These subtle fluctuations in the density of galaxy population can lead to ``cosmic variance'' \citep{driver2010quantifying,moster2011cosmic}, which can impact measurements at a cosmological scale \citep{kamionkowski1997getting,amarena2018impact,keenan2020biases}. 

The probe of asymmetry between galaxies spinning in opposite directions is a relative measurement rather than an absolute measurement. That is, the asymmetry is determined by the difference between two measurements made in the same field, and therefore should not be affected by cosmic variance. Any cosmic variance or other effects that impacts the number of clockwise galaxies observed from Earth is expected to have a similar effect on the number of counterclockwise galaxies.

\subsection{Multiple objects at the same galaxy}

In some cases, digital sky surveys can identify several photometric objects as independent galaxies, even in cases they are part of one larger galaxy. In the datasets used here all photometric objects that are part of the same galaxy were removed by removing all objects that had another object within 0.01$^o$.

Even if such objects existed in the datasets, they are expected to be distributed evenly between galaxies that spin clockwise and galaxies that spin counterclockwise, and therefore should not introduce an asymmetry. Experiments by using datasets of galaxies assigned with random spin directions and adding artificial objects to the galaxies showed that adding objects at exactly the same position of the original galaxies does not lead to signal of asymmetry \citep{shamir2021particles}. 

The experiments were made by using $\sim7.7\cdot10^4$ SDSS galaxies, and assigning the galaxies with random spin directions. Then, gradually adding more objects with the same location and spin directions as the galaxies in the original dataset, and the new artificial galaxies were assigned with the same spin direction as the galaxies in the original dataset \citep{shamir2021particles}. Adding such artificial galaxies did not lead to statistically significant signal.

\subsection{Differences in inclination}

The axis shown in Section~\ref{dipole} is profiled by computing the most likely axis when limiting the galaxies to different redshift ranges. In each redshift range, the axis is computed by the cosine dependence of all galaxies in the dataset that fit the redshift range, regardless of their location in the sky. By applying simple geometry, the change in the location of the most likely axis in different redshift ranges can be used to deduce the location of an axis that does not necessarily goes directly through Earth.

That analysis can also be affected by differences in the inclination of galaxies. The inclination of the galaxies can impact the ability to identify its spin direction, as a sharper inclination makes it more difficult to identify the spin patterns. As mentioned above, the axis is computed for each redshift range by fitting all galaxies in that redshift range, regardless of their position in the sky, and therefore it is expected that the impact of the inclination would be statistically the same among clockwise and counterclockwise galaxies. 

To test that assumption, the inclination of galaxies that spin clockwise was compared to the inclination of galaxies that spin counterclockwise. The inclination of each galaxy was computed by $\cos^{-1}(\frac{short \: axis}{long \: axis})$, and the short and long axis of all galaxies were determined by {\it Ganalyzer} as described in \citep{shamir2011ganalyzer}. Table~\ref{inclination} shows the average inclination of the galaxies in different redshift ranges and different RA ranges.

\begin{table}
\centering
\scriptsize
\begin{tabular}{llll}
\hline
Z           & RA         & Average     & Average     \\
range    & range              &  inclination                             &   inclination                             \\
            &      & cw (rad)                   & ccw (rad)                \\
\hline
$<0.15$   & all     &  1.14372$\pm$0.001 &   1.14426$\pm$0.001 \\
$>0.15$  & all      &  1.18472$\pm$0.002 &   1.18529$\pm$0.002  \\

$<0.15$   & $(0^o<\alpha<180^o)$        &  1.14387$\pm$0.002   & 1.14459$\pm$0.002 \\
$<0.15$   & $(180^o<\alpha<360^o)$     &  1.14357$\pm$0.002 & 1.14371$\pm$0.002 \\

$>0.15$   & $(0^o<\alpha<180^o)$         &  1.18488$\pm$0.003 & 1.18564$\pm$0.003 \\
$>0.15$   & $(180^o<\alpha<360^o)$     &  1.18451$\pm$0.002 & 1.18464$\pm$0.002 \\

\hline
\end{tabular}
\caption{Average inclination (in radians) of clockwise galaxies and counterclockwise galaxies in different redshift and RA ranges. The errors are the standard errors of the means.}
\label{inclination}
\end{table}

As expected, the table shows that the difference in inclination between clockwise and counterclockwise galaxies is not statistically significant, and well below the standard error. The inclination in higher redshifts is somewhat higher than the lower redshift, and that difference can be attributed to the smaller size of the galaxies. But also in the higher redshifts, there are no statistically significant differences between the inclination of clockwise galaxies and the inclination of counterclockwise galaxies. As also expected, no differences in the inclination were observed in opposite hemispheres.

\subsection{Photometric redshift}

Some of the analysis shown here is based on the redshift of the galaxies. Obtaining the spectra of a galaxy is a relatively long process, and therefore the vast majority of the galaxies do not have spectra. To estimate the redshift of galaxies that do not have spectra, the redshift can be estimated computationally from the photometric information in an approach called ``photometric redshift". While the photometric redshift is very quick to compute, it is also highly inaccurate, ambigous (meaning that one galaxy can have several different photometric redshifts), and systematically biased. 

In this study the asymmetry is of magnitude smaller than 1\%. The error of the state-of-the-art photometric redshift methods is 10\% to 20\%. Since it is normally determined by complex data-driven rules of machine learning systems, it is also systematically bias in a manner that is difficult to quantify and profile. Therefore, the photometric redshift is not a suitable probe that can be used in this study. All redshifts used in this paper are the spectroscopic redshifts, and the photometric redshifts are not used in any part of this study. Some of the analyses are done with no redshift information at all, showing that the signal is not originated from an error or systematic bias in the redshift.

\subsection{Bias carried over from previous catalogs}
\label{morphology_catalogs}

Catalogs of galaxy morphology can be prepared by either manual annotation of the galaxies \citep{land2008galaxy,nair2010catalog,baillard2011efigi}, or by automatic annotation \citep{gravet2015catalog,perez2015morphologies,goddard2020catalog,cheng2021galaxy}. When not prepared specifically for the purpose of analysis of galaxy spin directions, such catalogs can be biased in some way by the process through which the galaxies were annotated. For instance, \cite{land2008galaxy} found very substantial bias in the annotation of galaxies by their spin direction when the annotation was done by anonymous volunteers. Such bias driven by the human perception is difficult to quantify and correct, and even a small but consistent bias can lead to strong signal in the analysis \citep{shamir2021particles}. Such bias can also affect the separation of galaxies into elliptical and spiral galaxies, and therefore using a catalog that was prepared manually could lead to unexpected patterns that might be driven by the human perception rather than the real sky.

Some catalogs of galaxy morphology were prepared by using certain algorithms. Examples of such catalogs include \citep{gravet2015catalog,perez2015morphologies,goddard2020catalog,cheng2021galaxy}. However, these catalogs rely in most cases on machine learning algorithms, which work by complex data-driven rules. Due to their complexity and non-intuitive nature, it is very difficult to verify that these algorithms are fully symmetric. More importantly, these machine learning systems are based on manual annotation of the galaxies, and therefore any bias in the manual annotation would be carried on to the catalog. For instance, \citep{goddard2020catalog,cheng2021galaxy} are two catalogs that made use of crowdsourcing annotations that are known to be biased \citep{land2008galaxy,hayes2017nature}, and therefore are not safe for the task shown in this study, any other task related to an analysis of anisotropy in the large-scale structure. Moreover, algorithms based on deep neural networks use any discriminative information they can find in the classes of images, and therefore also learn the background of the images, leading to unusual and unexpected biases that are difficult to profile, as explained in detail in \cite{dhar2022systematic}.

An important aspect of the experimental design of this study is that no catalog of galaxy morphology was used. The entire process of annotation was done by the model-driven symmetric algorithm explained in Section~\ref{ganalyzer}, and no assumptions are made about the distribution of the galaxies in any existing catalog of galaxy morphology. The only selection of the galaxies is by the object detection algorithms and the magnitude of the objects in HST, DES, DESI Legacy Survey, SDSS, and Pan-STARRS photometric pipelines. It is difficult to think of a bias in these algorithms that would prefer one spin direction over another, and would also inverse that preference in the opposite hemisphere. While it is difficult to think of such bias in one sky survey, it is highly unlikely that such bias would appear in several sky surveys, with consistent profiles such that the profile of the bias in one sky survey matches the profile of the bias in the other surveys.

\subsection{Atmospheric effect}

There is no known atmospheric effect that can make a galaxy that spin clockwise appear as if it spins counterclockwise. Also, because the asymmetry is always measured with galaxies imaged in the same field, any kind of atmospheric effect that affects galaxies the spin clockwise will also affect galaxies that spin counterclockwise. Therefore, it is unlikely that a certain atmospheric effect would impact the number of clockwise galaxies at a certain field, but would have different impact on galaxies spinning counterclockwise. In any case, one of the datasets used here is made of galaxies imaged by the space-based HST, and are therefore not subjected to any kind of atmospheric effect.

\subsection{Backward spiral galaxies}

In rare cases, the shape of the arms of a spiral galaxy is not an indication of the spin direction of the galaxy. An example is NGC 4622 \citep{freeman1991simulating,buta2003ringed}. A prevalent and systematically uneven distribution of backward spiral galaxies might indeed lead to asymmetry between the number of galaxies spinning clockwise and the number of galaxies spinning counterclockwise. For instance, if a relatively high percentage of galaxies that actually spin clockwise are backward spiral galaxies, it would have led to an excessive number of galaxies that seem to be spinning counterclockwise.

However, backward spiral galaxies are relatively rare. Also, these galaxies are expected to be distributed equally between galaxies that spin clockwise and galaxies that spin counterclockwise, and there is no indication of asymmetry between backwards spiral galaxies. Therefore, according to the known evidence, there is no reason to assume that the observations shown here are driven by backward spiral galaxies.

\section{Conclusion}
\label{conclusion}

Autonomous digital sky surveys powered by robotic telescopes have allowed the collection of unprecedented amounts of astronomical data, enabling to address research questions that their studying was not feasible in the pre-information era. For instance, very large structures such as the Great Wall of Sloan were not discovered until robotic telescopes that can collect very large databases were introduced.

The question addressed here is the large-scale distribution of the spin directions of spiral galaxies. Multiple previous experiments have shown that the distribution of spin directions of spiral galaxies as observed from Earth might not be random \citep{longo2011detection,shamir2012handedness,shamir2013color,shamir2016asymmetry,shamir2017large,shamir2017photometric,shamir2017colour,shamir2019large,shamir2020pasa,shamir2020patterns,lee2019galaxy,lee2019mysterious,shamir2021particles,shamir2021large}.  

This study uses a large dataset of galaxies with spectra from several different telescopes. The analysis shows that the asymmetry in the spin directions of spiral galaxies increases with the redshift. The peak of the most likely axis changes consistently with the redshift, which can imply on an axis that does not go directly through Earth. The analysis uses several telescopes, covering the both the Northern and Southern hemispheres. The findings are in agreement with previous results, including space-based data acquired by the Hubble Space Telescope. Another noted observation is that the asymmetry becomes stronger as the redshift gets higher. Although the redshift range of the galaxies used in this study is naturally limited by the imaging capabilities of the telescopes, that correlation can be interpreted as higher asymmetry in the earlier Universe. If that trend is consistent also in higher redshifts not observed in this study, it can be viewed as an indication that the asymmetry patterns are primordial, and were stronger in the young Universe, but gradually becomes weaker as the Universe gets older. That can be explained by gravitational interaction between galaxies and galaxy mergers that can change the spin direction of the galaxies. Future sky surveys such as the Vera Rubin Observatory will have far greater depth, and will allow to test whether the trend continues also in higher redshifts.

Studies with smaller datasets of galaxies showed non-random spin directions of galaxies in filaments of the cosmic web, as described in \citep{tempel2013evidence,tempel2013galaxy,tempel2014detecting,dubois2014dancing,kraljic2021sdss} and others. Other studies showed alignment in the spin directions even when the galaxies are too far from each other to interact gravitationally \citep{lee2019galaxy,lee2019mysterious}, unless assuming modified Newtonian dynamics (MOND) gravity models that explain longer gravitational span \citep{sanders2003missing,darabi2014liquid,amendola2020measuring}.

Other observations of large-scale alignment in spin directions focused on quasars such as \citep{hutsemekers1998evidence,hutsemekers2005mapping,agarwal2011alignments,hutsemekers2014alignment}. Position angle of radio galaxies also showed large-scale consistency of angular momentum \citep{taylor2016alignments}. These observations agree with observations made with datasets such as the Faint Images of the Radio Sky at Twenty-centimetres (FIRST) and the TIFR GMRT Sky Survey (TGSS), showing large-scale alignment of radio galaxies \citep{contigiani2017radio,panwar2020alignment}.

These observational studies are also supported by simulations of dark matter \citep{aragon2007spin,zhang2009spin,codis2012connecting,libeskind2013velocity,libeskind2014universal,ganeshaiah2018cosmic,kraljic2020and} and galaxies \citep{dubois2014dancing,codis2018galaxy,ganeshaiah2019cosmic,kraljic2020and}, showing links between spin directions and the large-scale structure. That correlation was associated with the stellar mass and color of the galaxies \citep{wang2018spin}, and it has been proposed that the association was also linked to halo formation \citep{wang2017general}. That led to the contention that the spin direction in the halo progenitors is linked to the large-scale structure of the Universe \citep{wang2018build}. It should be mentioned that the spin direction of a galaxy might not be necessarily the same as the spin direction of the dark matter halo, as it has been proposed that in some cases a galaxy might spin in a different direction than its host dark matter halo \citep{wang2018spin}.

The analysis of spin directions done in this study provides evidence of large-scale dipole and quadrupole alignment. The observation of a large-scale axis has been proposed in the past by analyzing the cosmic microwave background (CMB), with consistent data from the Cosmic Background Explorer (COBE), Wilkinson Microwave Anisotropy Probe (WMAP) and Planck, as described in  \citep{abramo2006anomalies,mariano2013cmb,land2005examination,ade2014planck,santos2015influence,gruppuso2018evens} and other studies. Observations also showed that the axis formed by the CMB temperature is aligned with other cosmic asymmetry axes such as dark energy and dark flow \citep{mariano2013cmb}. Other notable statistical anomalies in the CMB are the quandrupole-octopole alignment \citep{schwarz2004low,ralston2004virgo,copi2007uncorrelated,copi2010large,copi2015large}, the asymmetry between hemispheres \citep{eriksen2004asymmetries,land2005examination,akrami2014power}, point-parity asymmetry \citep{kim2010anomalous,kim2010anomalous2}, and the CMB Cold Spot. If these anomalies are not statistical fluctuations \citep{bennett2011seven}, they can be viewed as observations that disagree with $\Lambda$CDM, as proposed by \cite{bull2016beyond,yeung2022directional} and others.

As described in Section~\ref{introduction}, the possible axis observed in the CMB is aligned with theories related to the geometry of the Universe such as ellipsoidal universe \citep{campanelli2006ellipsoidal,campanelli2007cosmic,gruppuso2007complete,rodrigues2008anisotropic,campanelli2011cosmic,cea2014ellipsoidal}, rotating universe \citep{gamow1946rotating,godel1949example,ozsvath1962finite,ozsvath2001approaches,su2009universe,sivaram2012primordial,chechin2016rotation,chechin2017does,camp2021}, holographic big bang \citep{pourhasan2014out,altamirano2017cosmological}, negative gravitational mass \citep{le2021recent}, and Black hole cosmology \citep{pathria1972universe,stuckey1994observable,easson2001universe,seshavatharam2010physics,poplawski2010radial,poplawski2010cosmology,chakrabarty2020toy,seshavatharam2020light,rinaldi2022matrix}

The availability of robotic telescopes provides the ability to analyze a possible non-random distribution of the spin directions of spiral galaxies, and that research question was not approachable in the pre-information era. As evidence for such non-random distribution are accumulating, additional research will be needed to fully understand its nature, and match it with other messengers in addition to CMB.

\section*{Acknowledgments}

I would like to thank the anonymous reviewer for the insightful comments. This study was supported in part by NSF grants AST-1903823 and IIS-1546079. I would like to thank Ethan Nguyen for retrieving and organizing the image data from the DESI Legacy Survey. 

This project used public archival data from the Dark Energy Survey (DES). Funding for the DES Projects has been provided by the U.S. Department of Energy, the U.S. National Science Foundation, the Ministry of Science and Education of Spain, the Science and Technology FacilitiesCouncil of the United Kingdom, the Higher Education Funding Council for England, the National Center for Supercomputing Applications at the University of Illinois at Urbana-Champaign, the Kavli Institute of Cosmological Physics at the University of Chicago, the Center for Cosmology and Astro-Particle Physics at the Ohio State University, the Mitchell Institute for Fundamental Physics and Astronomy at Texas A\&M University, Financiadora de Estudos e Projetos, Funda{\c c}{\~a}o Carlos Chagas Filho de Amparo {\`a} Pesquisa do Estado do Rio de Janeiro, Conselho Nacional de Desenvolvimento Cient{\'i}fico e Tecnol{\'o}gico and the Minist{\'e}rio da Ci{\^e}ncia, Tecnologia e Inova{\c c}{\~a}o, the Deutsche Forschungsgemeinschaft, and the Collaborating Institutions in the Dark Energy Survey.

The Collaborating Institutions are Argonne National Laboratory, the University of California at Santa Cruz, the University of Cambridge, Centro de Investigaciones Energ{\'e}ticas, Medioambientales y Tecnol{\'o}gicas-Madrid, the University of Chicago, University College London, the DES-Brazil Consortium, the University of Edinburgh, the Eidgen{\"o}ssische Technische Hochschule (ETH) Z{\"u}rich,  Fermi National Accelerator Laboratory, the University of Illinois at Urbana-Champaign, the Institut de Ci{\`e}ncies de l'Espai (IEEC/CSIC), the Institut de F{\'i}sica d'Altes Energies, Lawrence Berkeley National Laboratory, the Ludwig-Maximilians Universit{\"a}t M{\"u}nchen and the associated Excellence Cluster Universe, the University of Michigan, the National Optical Astronomy Observatory, the University of Nottingham, The Ohio State University, the OzDES Membership Consortium, the University of Pennsylvania, the University of Portsmouth, SLAC National Accelerator Laboratory, Stanford University, the University of Sussex, and Texas A\&M University.

Based in part on observations at Cerro Tololo Inter-American Observatory, National Optical Astronomy Observatory, which is operated by the Association of Universities for Research in Astronomy (AURA) under a cooperative agreement with the National Science Foundation.

The research is based on observations made with the NASA/ESA Hubble Space Telescope, and obtained from the Hubble Legacy Archive, which is a collaboration between the Space Telescope Science Institute (STScI/NASA), the Space Telescope European Coordinating Facility (ST-ECF/ESA) and the Canadian Astronomy Data Centre (CADC/NRC/CSA).

The Legacy Surveys consist of three individual and complementary projects: the Dark Energy Camera Legacy Survey (DECaLS; Proposal ID \#2014B-0404; PIs: David Schlegel and Arjun Dey), the Beijing-Arizona Sky Survey (BASS; NOAO Prop. ID \#2015A-0801; PIs: Zhou Xu and Xiaohui Fan), and the Mayall z-band Legacy Survey (MzLS; Prop. ID \#2016A-0453; PI: Arjun Dey). DECaLS, BASS and MzLS together include data obtained, respectively, at the Blanco telescope, Cerro Tololo Inter-American Observatory, NSF’s NOIRLab; the Bok telescope, Steward Observatory, University of Arizona; and the Mayall telescope, Kitt Peak National Observatory, NOIRLab. The Legacy Surveys project is honored to be permitted to conduct astronomical research on Iolkam Du’ag (Kitt Peak), a mountain with particular significance to the Tohono O’odham Nation.

NOIRLab is operated by the Association of Universities for Research in Astronomy (AURA) under a cooperative agreement with the National Science Foundation.

This project used data obtained with the Dark Energy Camera (DECam), which was constructed by the Dark Energy Survey (DES) collaboration. Funding for the DES Projects has been provided by the U.S. Department of Energy, the U.S. National Science Foundation, the Ministry of Science and Education of Spain, the Science and Technology Facilities Council of the United Kingdom, the Higher Education Funding Council for England, the National Center for Supercomputing Applications at the University of Illinois at Urbana-Champaign, the Kavli Institute of Cosmological Physics at the University of Chicago, Center for Cosmology and Astro-Particle Physics at the Ohio State University, the Mitchell Institute for Fundamental Physics and Astronomy at Texas A\&M University, Financiadora de Estudos e Projetos, Fundacao Carlos Chagas Filho de Amparo, Financiadora de Estudos e Projetos, Fundacao Carlos Chagas Filho de Amparo a Pesquisa do Estado do Rio de Janeiro, Conselho Nacional de Desenvolvimento Cientifico e Tecnologico and the Ministerio da Ciencia, Tecnologia e Inovacao, the Deutsche Forschungsgemeinschaft and the Collaborating Institutions in the Dark Energy Survey. The Collaborating Institutions are Argonne National Laboratory, the University of California at Santa Cruz, the University of Cambridge, Centro de Investigaciones Energeticas, Medioambientales y Tecnologicas-Madrid, the University of Chicago, University College London, the DES-Brazil Consortium, the University of Edinburgh, the Eidgenossische Technische Hochschule (ETH) Zurich, Fermi National Accelerator Laboratory, the University of Illinois at Urbana-Champaign, the Institut de Ciencies de l’Espai (IEEC/CSIC), the Institut de Fisica d’Altes Energies, Lawrence Berkeley National Laboratory, the Ludwig Maximilians Universitat Munchen and the associated Excellence Cluster Universe, the University of Michigan, NSF’s NOIRLab, the University of Nottingham, the Ohio State University, the University of Pennsylvania, the University of Portsmouth, SLAC National Accelerator Laboratory, Stanford University, the University of Sussex, and Texas A\&M University.

BASS is a key project of the Telescope Access Program (TAP), which has been funded by the National Astronomical Observatories of China, the Chinese Academy of Sciences (the Strategic Priority Research Program “The Emergence of Cosmological Structures” Grant \# XDB09000000), and the Special Fund for Astronomy from the Ministry of Finance. The BASS is also supported by the External Cooperation Program of Chinese Academy of Sciences (Grant \# 114A11KYSB20160057), and Chinese National Natural Science Foundation (Grant \# 11433005).

The Legacy Survey team makes use of data products from the Near-Earth Object Wide-field Infrared Survey Explorer (NEOWISE), which is a project of the Jet Propulsion Laboratory/California Institute of Technology. NEOWISE is funded by the National Aeronautics and Space Administration.

The Legacy Surveys imaging of the DESI footprint is supported by the Director, Office of Science, Office of High Energy Physics of the U.S. Department of Energy under Contract No. DE-AC02-05CH1123, by the National Energy Research Scientific Computing Center, a DOE Office of Science User Facility under the same contract; and by the U.S. National Science Foundation, Division of Astronomical Sciences under Contract No. AST-0950945 to NOAO.

SDSS-IV is managed by the Astrophysical Research Consortium for the Participating Institutions of the SDSS Collaboration including the Brazilian Participation Group, the Carnegie Institution for Science, Carnegie Mellon University, the Chilean Participation Group, the French Participation Group, Harvard-Smithsonian Center for Astrophysics, Instituto de Astrofisica de Canarias, The Johns Hopkins University, Kavli Institute for the Physics and Mathematics of the Universe (IPMU) / University of Tokyo, the Korean Participation Group, Lawrence Berkeley National Laboratory, Leibniz Institut fur Astrophysik Potsdam (AIP), Max-Planck-Institut fur Astronomie (MPIA Heidelberg), Max-Planck-Institut fur Astrophysik (MPA Garching), Max-Planck-Institut fur Extraterrestrische Physik (MPE), National Astronomical Observatories of China, New Mexico State University, New York University, University of Notre Dame, Observatario Nacional / MCTI, The Ohio State University, Pennsylvania State University, Shanghai Astronomical Observatory, United Kingdom Participation Group, Universidad Nacional Autonoma de Mexico, University of Arizona, University of Colorado Boulder, University of Oxford, University of Portsmouth, University of Utah, University of Virginia, University of Washington, University of Wisconsin, Vanderbilt University, and Yale University.

The Pan-STARRS1 Surveys (PS1) and the PS1 public science archive have been made possible through contributions by the Institute for Astronomy, the University of Hawaii, the Pan-STARRS Project Office, the Max-Planck Society and its participating institutes, the Max Planck Institute for Astronomy, Heidelberg and the Max Planck Institute for Extraterrestrial Physics, Garching, The Johns Hopkins University, Durham University, the University of Edinburgh, the Queen's University Belfast, the Harvard-Smithsonian Center for Astrophysics, the Las Cumbres Observatory Global Telescope Network Incorporated, the National Central University of Taiwan, the Space Telescope Science Institute, the National Aeronautics and Space Administration under Grant No. NNX08AR22G issued through the Planetary Science Division of the NASA Science Mission Directorate, the National Science Foundation Grant No. AST-1238877, the University of Maryland, Eotvos Lorand University (ELTE), the Los Alamos National Laboratory, and the Gordon and Betty Moore Foundation.

\bibliographystyle{apalike}
\bibliography{main_arxiv}

\begin{thebibliography}{}

\bibitem[Aab et~al., 2017]{aab2017observation}
Aab, A., Abreu, P., Aglietta, M., Al~Samarai, I., Albuquerque, I., Allekotte,
  I., Almela, A., Castillo, J.~A., Alvarez-Mu{\~n}iz, J., anastasi, G.~A.,
  et~al. (2017).
\newblock Observation of a large-scale anisotropy in the arrival directions of
  cosmic rays above 8$\times$ 1018 ev.
\newblock {\em Science}, 357(6357):1266--1270.

\bibitem[Abramo et~al., 2006]{abramo2006anomalies}
Abramo, L.~R., Sodr{\'e}~Jr, L., and Wuensche, C.~A. (2006).
\newblock anomalies in the low cmb multipoles and extended foregrounds.
\newblock {\em Physical Review D}, 74(8):083515.

\bibitem[Ade et~al., 2014]{ade2014planck}
Ade, P.~A., Aghanim, N., Armitage-Caplan, C., Arnaud, M., Ashdown, M.,
  Atrio-Barandela, F., Aumont, J., Baccigalupi, C., Banday, a.~J., Barreiro,
  R., et~al. (2014).
\newblock Planck 2013 results. xxiii. isotropy and statistics of the cmb.
\newblock {\em Astronomy and Astrophysics}, 571:A23.

\bibitem[Adelman-McCarthy, 2008]{adelman2008ma}
Adelman-McCarthy, J. (2008).
\newblock Ma ag ueros, ss allam, et al.
\newblock {\em Astrophysical Journal Supplement Series}, 175:297.

\bibitem[Adhav, 2011]{adhav2011lrs}
Adhav, K. (2011).
\newblock Lrs bianchi type-i universe with anisotropic dark energy in lyra
  geometry.
\newblock {\em International Journal of Astronomy and Astrophysics},
  1(4):204--209.

\bibitem[Adhav et~al., 2011]{adhav2011kantowski}
Adhav, K., Bansod, A., Wankhade, R.~J., and Astronomical~Journalmire, H.
  (2011).
\newblock Kantowski-sachs cosmological models with anisotropic dark energy.
\newblock {\em Open Physics}, 9(4):919--925.

\bibitem[Agarwal et~al., 2011]{agarwal2011alignments}
Agarwal, N., Kamal, A., and Jain, P.~J. (2011).
\newblock Alignments in quasar polarizations: pseudoscalar-photon mixing in the
  presence of correlated magnetic fields.
\newblock {\em Physical Review D}, 83(6):065014.

\bibitem[Akrami et~al., 2014]{akrami2014power}
Akrami, Y., Fantaye, Y., Shafieloo, A., Eriksen, H., Hansen, F.~K., Banday,
  a.~J., and G{\'o}rski, K.~M. (2014).
\newblock Power asymmetry in wmap and planck temperature sky maps as measured
  by a local variance estimator.
\newblock {\em Astrophysical Journal Letters}, 784(2):L42.

\bibitem[Altamirano et~al., 2017]{altamirano2017cosmological}
Altamirano, N., Gould, E., Afshordi, N., and Mann, R.~B. (2017).
\newblock Cosmological perturbations in the 5d holographic big bang model.
\newblock {\em arXiv preprint arXiv:1703.00954}.

\bibitem[Amendola et~al., 2020]{amendola2020measuring}
Amendola, L., Bettoni, D., Pinho, a.~M., and Casas, S. (2020).
\newblock Measuring gravity at cosmological scales.
\newblock {\em Universe}, 6(2):20.

\bibitem[Arag{\'o}n-Calvo et~al., 2007]{aragon2007spin}
Arag{\'o}n-Calvo, M.~A., van~de Weygaert, R., Jones, B.~J., and Van Der~Hulst,
  J. (2007).
\newblock Spin alignment of dark matter halos in filaments and walls.
\newblock {\em Astrophysical Journal}, 655(1):L5.

\bibitem[Arciniega et~al., 2020a]{arciniega2020geometric}
Arciniega, G., Bueno, P., Cano, P.~A., Edelstein, J.~D., Hennigar, R.~A., and
  Jaime, L.~G. (2020a).
\newblock Geometric inflation.
\newblock {\em Physics Letters B}, 802:135242.

\bibitem[Arciniega et~al., 2020b]{arciniega2020towards}
Arciniega, G., Edelstein, J.~D., and Jaime, L.~G. (2020b).
\newblock Towards geometric inflation: the cubic case.
\newblock {\em Physics Letters B}, 802:135272.

\bibitem[Azarnia et~al., 2021]{azarnia2021islands}
Azarnia, S., Fareghbal, R., Naseh, A., and Zolfi, H. (2021).
\newblock Islands in flat-space cosmology.
\newblock {\em Physical Review D}, 104(12):126017.

\bibitem[Baillard et~al., 2011]{baillard2011efigi}
Baillard, a., Bertin, E., De~Lapparent, V., Fouqu{\'e}, P., Arnouts, S.,
  Mellier, Y., Pell{\'o}, R., Leborgne, J.-F., Prugniel, P., Makarov, D.,
  et~al. (2011).
\newblock The efigi catalogue of 4458 nearby galaxies with detailed morphology.
\newblock {\em Astronomy and Astrophysics}, 532:A74.

\bibitem[Bak and Rey, 2000]{bak2000holographic}
Bak, D. and Rey, S.-J. (2000).
\newblock Holographic principle and string cosmology.
\newblock {\em Classical and Quantum Gravity}, 17(1):L1.

\bibitem[Beltran~Jimenez and Maroto, 2007]{jimenez2007cosmology}
Beltran~Jimenez, J. and Maroto, a.~L. (2007).
\newblock Cosmology with moving dark energy and the cmb quadrupole.
\newblock {\em Physical Review D}, 76(2):023003.

\bibitem[Bennett et~al., 2011]{bennett2011seven}
Bennett, C., Hill, R., Hinshaw, G., Larson, D., Smith, K., Dunkley, J., Gold,
  B., Halpern, M., Jarosik, N., Kogut, A., et~al. (2011).
\newblock Seven-year wilkinson microwave anisotropy probe (wmap*) observations:
  Are there cosmic microwave background anomalies?
\newblock {\em Astrophysical Journal Supplement Series}, 192(2):17.

\bibitem[Bohmer and Mota, 2008]{bohmer2008cmb}
Bohmer, C.~G. and Mota, D.~F. (2008).
\newblock Cmb anisotropies and inflation from non-standard spinors.
\newblock {\em Physics Letters B}, 663(3):168--171.

\bibitem[Bousso, 2002]{bousso2002holographic}
Bousso, R. (2002).
\newblock The holographic principle.
\newblock {\em Reviews of Modern Physics}, 74(3):825.

\bibitem[Bull et~al., 2016]{bull2016beyond}
Bull, P., Akrami, Y., Adamek, J., Baker, T., Bellini, E., Jimenez, J.~B.,
  Bentivegna, E., Camera, S., Clesse, S., Davis, J.~H., et~al. (2016).
\newblock Beyond $\lambda$cdm: Problems, solutions, and the road ahead.
\newblock {\em Physics of the Dark Universe}, 12:56--99.

\bibitem[Buta et~al., 2003]{buta2003ringed}
Buta, R.~J., Byrd, G.~G., and Freeman, T. (2003).
\newblock The ringed spiral galaxy ngc 4622. i. photometry, kinematics, and the
  case for two strong leading outer spiral arms.
\newblock {\em Astronomical Journal}, 125(2):634.

\bibitem[Camarena and Marra, 2018]{amarena2018impact}
Camarena, D. and Marra, V. (2018).
\newblock Impact of the cosmic variance on h 0 on cosmological analyses.
\newblock {\em Physical Review D}, 98(2):023537.

\bibitem[Campanelli, 2021]{camp2021}
Campanelli, L. (2021).
\newblock A conjecture on the neutrality of matter.
\newblock {\em Foundations of Physics}, 51:56.

\bibitem[Campanelli et~al., 2011]{campanelli2011cosmic}
Campanelli, L., Cea, P., Fogli, G., and Tedesco, L. (2011).
\newblock Cosmic parallax in ellipsoidal universe.
\newblock {\em Modern Physics Letters A}, 26(16):1169--1181.

\bibitem[Campanelli et~al., 2006]{campanelli2006ellipsoidal}
Campanelli, L., Cea, P., and Tedesco, L. (2006).
\newblock Ellipsoidal universe can solve the cosmic microwave background
  quadrupole problem.
\newblock {\em Physics Review Letters}, 97(13):131302.

\bibitem[Campanelli et~al., 2007]{campanelli2007cosmic}
Campanelli, L., Cea, P., and Tedesco, L. (2007).
\newblock Cosmic microwave background quadrupole and ellipsoidal universe.
\newblock {\em Physical Review D}, 76(6):063007.

\bibitem[Carter et~al., 2020]{carter2020overinterpretation}
Carter, B., Jain, S., Mueller, J., and Gifford, D. (2020).
\newblock Overinterpretation reveals image classification model pathologies.
\newblock {\em arXiv:2003.08907}.

\bibitem[Cea, 2014]{cea2014ellipsoidal}
Cea, P. (2014).
\newblock The ellipsoidal universe in the planck satellite era.
\newblock {\em Monthly Notices of the Royal Astronomical Society},
  441(2):1646--1661.

\bibitem[Chakrabarty et~al., 2020]{chakrabarty2020toy}
Chakrabarty, H., Abdujabbarov, A., Malafarina, D., and Bambi, C. (2020).
\newblock A toy model for a baby universe inside a black hole.
\newblock {\em European Physical Journal C}, 80(1909.07129):1--10.

\bibitem[Chechin, 2016]{chechin2016rotation}
Chechin, L. (2016).
\newblock Rotation of the universe at different cosmological epochs.
\newblock {\em Astronomy Reports}, 60(6):535--541.

\bibitem[Chechin, 2017]{chechin2017does}
Chechin, L. (2017).
\newblock Does the cosmological principle exist in the rotating universe?
\newblock {\em Gravitation and Cosmology}, 23(4):305--310.

\bibitem[Cheng et~al., 2021]{cheng2021galaxy}
Cheng, T.-Y., Conselice, C.~J., Arag{\'o}n-Salamanca, A., Aguena, M., Allam,
  S., andrade Oliveira, F., annis, J., Bluck, A., Brooks, D., Burke, D., et~al.
  (2021).
\newblock Galaxy morphological classification catalogue of the dark energy
  survey year 3 data with convolutional neural networks.
\newblock {\em Monthly Notices of the Royal Astronomical Society},
  507(3):4425--4444.

\bibitem[Christillin, 2014]{christillin2014machian}
Christillin, P. (2014).
\newblock The machian origin of linear inertial forces from our gravitationally
  radiating black hole universe.
\newblock {\em The European Physical Journal Plus}, 129(8):1--3.

\bibitem[Cline et~al., 2003]{cline2003does}
Cline, J.~M., Crotty, P., and Lesgourgues, J. (2003).
\newblock Does the small cmb quadrupole moment suggest new physics?
\newblock {\em Journal of Cosmology and Astroparticle Physics}, 2003(09):010.

\bibitem[Codis et~al., 2018]{codis2018galaxy}
Codis, S., Jindal, A., Chisari, N., Vibert, D., Dubois, Y., Pichon, C., and
  Devriendt, J. (2018).
\newblock Galaxy orientation with the cosmic web across cosmic time.
\newblock {\em Monthly Notices of the Royal Astronomical Society},
  481(4):4753--4774.

\bibitem[Codis et~al., 2012]{codis2012connecting}
Codis, S., Pichon, C., Devriendt, J., Slyz, A., Pogosyan, D., Dubois, Y., and
  Sousbie, T. (2012).
\newblock Connecting the cosmic web to the spin of dark haloes: implications
  for galaxy formation.
\newblock {\em Monthly Notices of the Royal Astronomical Society},
  427(4):3320--3336.

\bibitem[Cole et~al., 2005]{cole20052df}
Cole, S., Percival, W.~J., Peacock, J.~A., Norberg, P., Baugh, C.~M., Frenk,
  C.~S., Baldry, I., Bland-Hawthorn, J., Bridges, T., Cannon, R., et~al.
  (2005).
\newblock The 2df galaxy redshift survey: power-spectrum analysis of the final
  data set and cosmological implications.
\newblock {\em Monthly Notices of the Royal Astronomical Society},
  362(2):505--534.

\bibitem[Colin et~al., 2019]{colin2019evidence}
Colin, J., Mohayaee, R., Rameez, M., and Sarkar, S. (2019).
\newblock Evidence for anisotropy of cosmic acceleration.
\newblock {\em Astronomy and Astrophysics}, 631:L13.

\bibitem[Colless et~al., 2003]{colless20032df}
Colless, M., Peterson, B.~A., Jackson, C., Peacock, J.~A., Cole, S., Norberg,
  P., Baldry, I.~K., Baugh, C.~M., Bland-Hawthorn, J., and Bridges, T. A.~o.
  (2003).
\newblock The 2df galaxy redshift survey: final data release.
\newblock {\em arXiv preprint astro-ph/0306581}.

\bibitem[Contigiani et~al., 2017]{contigiani2017radio}
Contigiani, O., de~Gasperin, F., Miley, G., Rudnick, L., andernach, H.,
  Banfield, J., Kapi{\'n}ska, A., Shabala, S., and Wong, O. (2017).
\newblock Radio galaxy zoo: cosmological alignment of radio sources.
\newblock {\em Monthly Notices of the Royal Astronomical Society},
  472(1):636--646.

\bibitem[Copi et~al., 2007]{copi2007uncorrelated}
Copi, C.~J., Huterer, D., Schwarz, D.~J., and Starkman, G.~D. (2007).
\newblock Uncorrelated universe: statistical anisotropy and the vanishing
  angular correlation function in wmap years 1--3.
\newblock {\em Physical Review D}, 75(2):023507.

\bibitem[Copi et~al., 2010]{copi2010large}
Copi, C.~J., Huterer, D., Schwarz, D.~J., and Starkman, G.~D. (2010).
\newblock Large-angle anomalies in the cmb.
\newblock {\em Advances in Astronomy}, 2010.

\bibitem[Copi et~al., 2015]{copi2015large}
Copi, C.~J., Huterer, D., Schwarz, D.~J., and Starkman, G.~D. (2015).
\newblock Large-scale alignments from wmap and planck.
\newblock {\em Monthly Notices of the Royal Astronomical Society},
  449(4):3458--3470.

\bibitem[Darabi, 2014]{darabi2014liquid}
Darabi, F. (2014).
\newblock Liquid-droplet as a model for the rotation curve problem.
\newblock {\em Central European Journal of Physics}, 12(9):678--686.

\bibitem[De~Lapparent et~al., 1986]{de1986slice}
De~Lapparent, V., Geller, M.~J., and Huchra, J.~P. (1986).
\newblock A slice of the universe.
\newblock {\em Astrophysical Journal}, 302:L1--L5.

\bibitem[De~Vaucouleurs, 1958]{de1958tilt}
De~Vaucouleurs, G. (1958).
\newblock Tilt criteria and direction of rotation of spiral galaxies.
\newblock {\em Astrophysical Journal}, 127:487.

\bibitem[Deng et~al., 2006]{deng2006super}
Deng, X.-F., Chen, Y.-Q., Zhang, Q., and He, J.-Z. (2006).
\newblock Super-large-scale structures in the sloan digital sky survey.
\newblock {\em Chinese Journal of Astronomy and Astrophysics}, 6(1):35.

\bibitem[Dhar and Shamir, 2021]{dhar2021evaluation}
Dhar, S. and Shamir, L. (2021).
\newblock Evaluation of the benchmark datasets for testing the efficacy of deep
  convolutional neural networks.
\newblock {\em Visual Informatics}, 5(3):92--101.

\bibitem[Dhar and Shamir, 2022]{dhar2022systematic}
Dhar, S. and Shamir, L. (2022).
\newblock Systematic biases when using deep neural networks for annotating
  large catalogs of astronomical images.
\newblock {\em Astronomy and Computing}, 38:100545.

\bibitem[Dojcsak and Shamir, 2014]{dojcsak2014quantitative}
Dojcsak, L. and Shamir, L. (2014).
\newblock Quantitative analysis of spirality in elliptical galaxies.
\newblock {\em New Astronomy}, 28:1--8.

\bibitem[Dong et~al., 2015]{dong2015inflation}
Dong, Z., Ming-Hua, L., Ping, W., and Zhe, C. (2015).
\newblock Inflation in de sitter spacetime and cmb large scale anomaly.
\newblock {\em Chinese Physics C}, 39(9):095101.

\bibitem[Driver and Robotham, 2010]{driver2010quantifying}
Driver, S.~P. and Robotham, A.~S. (2010).
\newblock Quantifying cosmic variance.
\newblock {\em Monthly Notices of the Royal Astronomical Society},
  407(4):2131--2140.

\bibitem[Dubois et~al., 2014]{dubois2014dancing}
Dubois, Y., Pichon, C., Welker, C., Le~Borgne, D., Devriendt, J., Laigle, C.,
  Codis, S., Pogosyan, D., Arnouts, S., Benabed, K., et~al. (2014).
\newblock Dancing in the dark: galactic properties trace spin swings along the
  cosmic web.
\newblock {\em Monthly Notices of the Royal Astronomical Society},
  444(2):1453--1468.

\bibitem[Easson and Brandenberger, 2001]{easson2001universe}
Easson, D.~A. and Brandenberger, R.~H. (2001).
\newblock Universe generation from black hole interiors.
\newblock {\em Journal of High Energy Physics}, 2001(06):024.

\bibitem[Edelstein et~al., 2020]{edelstein2020aspects}
Edelstein, J.~D., Rodr{\'\i}guez, D.~V., and L{\'o}pez, A.~V. (2020).
\newblock Aspects of geometric inflation.
\newblock {\em Journal of Cosmology and Astroparticle Physics}, 2020(12):040.

\bibitem[Efstathiou, 2003]{efstathiou2003low}
Efstathiou, G. (2003).
\newblock Is the low cosmic microwave background quadrupole a signature of
  spatial curvature?
\newblock {\em Monthly Notices of the Royal Astronomical Society},
  343(4):L95--L98.

\bibitem[Eriksen et~al., 2004]{eriksen2004asymmetries}
Eriksen, H.~K., Hansen, F.~K., Banday, a.~J., Gorski, K.~M., and Lilje, P.~B.
  (2004).
\newblock Asymmetries in the cosmic microwave background anisotropy field.
\newblock {\em Astrophysical Journal}, 605(1):14.

\bibitem[Feng and Zhang, 2003]{feng2003double}
Feng, B. and Zhang, X. (2003).
\newblock Double inflation and the low cmb quadrupole.
\newblock {\em Physics Letters B}, 570(3-4):145--150.

\bibitem[Freeman et~al., 1991]{freeman1991simulating}
Freeman, T., Byrd, G., and Howard, S. (1991).
\newblock Simulating ngc 4622: A leading-arm spiral galaxy.
\newblock In {\em BASS}, volume~23, page 1460.

\bibitem[Gammie et~al., 2004]{gammie2004black}
Gammie, C.~F., Shapiro, S.~L., and McKinney, J.~C. (2004).
\newblock Black hole spin evolution.
\newblock {\em Astrophysical Journal}, 602(1):312.

\bibitem[Gamow, 1946]{gamow1946rotating}
Gamow, G. (1946).
\newblock Rotating universe?
\newblock {\em Nature}, 158(4016):549--549.

\bibitem[Ganeshaiah~Veena et~al., 2019]{ganeshaiah2019cosmic}
Ganeshaiah~Veena, P., Cautun, M., Tempel, E., van~de Weygaert, R., and Frenk,
  C.~S. (2019).
\newblock The cosmic ballet ii: spin alignment of galaxies and haloes with
  large-scale filaments in the eagle simulation.
\newblock {\em Monthly Notices of the Royal Astronomical Society},
  487(2):1607--1625.

\bibitem[Ganeshaiah~Veena et~al., 2018]{ganeshaiah2018cosmic}
Ganeshaiah~Veena, P., Cautun, M., van~de Weygaert, R., Tempel, E., Jones,
  B.~J., Rieder, S., and Frenk, C.~S. (2018).
\newblock The cosmic ballet: spin and shape alignments of haloes in the cosmic
  web.
\newblock {\em Monthly Notices of the Royal Astronomical Society},
  481(1):414--438.

\bibitem[Ghosh et~al., 2016]{ghosh2016probing}
Ghosh, S., Jain, P.~J., Kashyap, G., Kothari, R., Nadkarni-Ghosh, S., and
  Tiwari, P. (2016).
\newblock Probing statistical isotropy of cosmological radio sources using
  square kilometre array.
\newblock {\em Journal of Astrophysics and Astronomy}, 37(4):1--21.

\bibitem[Goddard and Shamir, 2020]{goddard2020catalog}
Goddard, H. and Shamir, L. (2020).
\newblock A catalog of broad morphology of pan-starrs galaxies based on deep
  learning.
\newblock {\em Astrophysical Journal Supplement Series}, 251(2):28.

\bibitem[G{\"o}del, 1949]{godel1949example}
G{\"o}del, K. (1949).
\newblock an example of a new type of cosmological solutions of einstein's
  field equations of gravitation.
\newblock {\em Reviews of Modern Physics}, 21(3):447.

\bibitem[Gordon and Hu, 2004]{gordon2004low}
Gordon, C. and Hu, W. (2004).
\newblock Low cmb quadrupole from dark energy isocurvature perturbations.
\newblock {\em Physical Review D}, 70(8):083003.

\bibitem[Gott~III et~al., 2005]{gott2005map}
Gott~III, J.~R., Juri{\'c}, M., Schlegel, D., Hoyle, F., Vogeley, M., Tegmark,
  M., Bahcall, N., and Brinkmann, J. (2005).
\newblock A map of the universe.
\newblock {\em Astrophysical Journal}, 624(2):463.

\bibitem[Gravet et~al., 2015]{gravet2015catalog}
Gravet, R., Cabrera-Vives, G., P{\'e}rez-Gonz{\'a}lez, P.~G., Kartaltepe, J.,
  Barro, G., Bernardi, M., Mei, S., Shankar, F., Dimauro, P., Bell, E., et~al.
  (2015).
\newblock A catalog of visual-like morphologies in the 5 candels fields using
  deep learning.
\newblock {\em Astrophysical Journal Supplement Series}, 221(1):8.

\bibitem[Grogin et~al., 2011]{grogin2011candels}
Grogin, N.~A., Kocevski, D.~D., Faber, S., Ferguson, H.~C., Koekemoer, a.~M.,
  Riess, A.~G., Acquaviva, V., Alexander, D.~M., Almaini, O., Ashby, M.~L.,
  et~al. (2011).
\newblock Candels: the cosmic assembly near-infrared deep extragalactic legacy
  survey.
\newblock {\em Astrophysical Journal Supplement Series}, 197(2):35.

\bibitem[Gruppuso, 2007]{gruppuso2007complete}
Gruppuso, A. (2007).
\newblock Complete statistical analysis for the quadrupole amplitude in an
  ellipsoidal universe.
\newblock {\em Physical Review D}, 76(8):083010.

\bibitem[Gruppuso et~al., 2018]{gruppuso2018evens}
Gruppuso, A., Kitazawa, N., Lattanzi, M., Mandolesi, N., Natoli, P., and
  Sagnotti, A. (2018).
\newblock The evens and odds of cmb anomalies.
\newblock {\em Physics of the Dark Universe}, 20:49--64.

\bibitem[Hawkins et~al., 2003]{hawkins20032df}
Hawkins, E., Maddox, S., Cole, S., Lahav, O., Madgwick, D.~S., Norberg, P.,
  Peacock, J.~A., Baldry, I.~K., Baugh, C.~M., Bland-Hawthorn, J., et~al.
  (2003).
\newblock The 2df galaxy redshift survey: correlation functions, peculiar
  velocities and the matter density of the universe.
\newblock {\em Monthly Notices of the Royal Astronomical Society},
  346(1):78--96.

\bibitem[Hayes et~al., 2017]{hayes2017nature}
Hayes, W.~B., Davis, D., and Silva, P. (2017).
\newblock On the nature and correction of the spurious s-wise spiral galaxy
  winding bias in galaxy zoo 1.
\newblock {\em Monthly Notices of the Royal Astronomical Society},
  466(4):3928--3936.

\bibitem[Hoehn and Shamir, 2014]{hoehn2014characteristics}
Hoehn, C. and Shamir, L. (2014).
\newblock Characteristics of clockwise and counterclockwise spiral galaxies.
\newblock {\em Astronomische Nachrichten}, 335(2):189--192.

\bibitem[Horv{\'a}th et~al., 2015]{horvath2015new}
Horv{\'a}th, I., Bagoly, Z., Hakkila, J., and T{\'o}th, L.~V. (2015).
\newblock New data support the existence of the hercules-corona borealis great
  wall.
\newblock {\em Astronomy and Astrophysics}, 584:A48.

\bibitem[Hu and Ling, 2006]{hu2006interacting}
Hu, B. and Ling, Y. (2006).
\newblock Interacting dark energy, holographic principle, and coincidence
  problem.
\newblock {\em Physical Review D}, 73(12):123510.

\bibitem[Hutsemekers, 1998]{hutsemekers1998evidence}
Hutsemekers, D. (1998).
\newblock Evidence for very large-scale coherent orientations of quasar
  polarization vectors.
\newblock {\em Astronomy and Astrophysics}, 332:410--428.

\bibitem[Hutsem{\'e}kers et~al., 2014]{hutsemekers2014alignment}
Hutsem{\'e}kers, D., Braibant, L., Pelgrims, V., and Sluse, D. (2014).
\newblock Alignment of quasar polarizations with large-scale structures.
\newblock {\em Astronomy and Astrophysics}, 572:A18.

\bibitem[Hutsem{\'e}kers et~al., 2005]{hutsemekers2005mapping}
Hutsem{\'e}kers, D., Cabanac, R., Lamy, H., and Sluse, D. (2005).
\newblock Mapping extreme-scale alignments of quasar polarization vectors.
\newblock {\em Astronomy and Astrophysics}, 441(3):915--930.

\bibitem[Jaime, 2021]{jaime2021viability}
Jaime, L.~G. (2021).
\newblock On the viability of the evolution of the universe with geometric
  inflation.
\newblock {\em Physics of the Dark Universe}, 34:100887.

\bibitem[Javanmardi and Kroupa, 2017]{javanmardi2017anisotropy}
Javanmardi, B. and Kroupa, P. (2017).
\newblock anisotropy in the all-sky distribution of galaxy morphological types.
\newblock {\em Astronomy and Astrophysics}, 597:A120.

\bibitem[Javanmardi et~al., 2015]{javanmardi2015probing}
Javanmardi, B., Porciani, C., Kroupa, P., and Pflam-Altenburg, J. (2015).
\newblock Probing the isotropy of cosmic acceleration traced by type ia
  supernovae.
\newblock {\em Astrophysical Journal}, 810(1):47.

\bibitem[Jones et~al., 2005]{jones2005scaling}
Jones, B.~J., Mart{\'\i}nez, V.~J., Saar, E., and Trimble, V. (2005).
\newblock Scaling laws in the distribution of galaxies.
\newblock {\em Reviews of Modern Physics}, 76(4):1211.

\bibitem[Kamionkowski and Loeb, 1997]{kamionkowski1997getting}
Kamionkowski, M. and Loeb, A. (1997).
\newblock Getting around cosmic variance.
\newblock {\em Physical Review D}, 56(8):4511.

\bibitem[Keenan et~al., 2020]{keenan2020biases}
Keenan, R.~P., Marrone, D.~P., and Keating, G.~K. (2020).
\newblock Biases and cosmic variance in molecular gas abundance measurements at
  high redshift.
\newblock {\em Astrophysical Journal}, 904(2):127.

\bibitem[Kim and Naselsky, 2010a]{kim2010anomalous2}
Kim, J. and Naselsky, P. (2010a).
\newblock anomalous parity asymmetry of the wilkinson microwave anisotropy
  probe power spectrum data at low multipoles.
\newblock {\em Astrophysical Journal Letters}, 714(2):L265.

\bibitem[Kim and Naselsky, 2010b]{kim2010anomalous}
Kim, J. and Naselsky, P. (2010b).
\newblock anomalous parity asymmetry of wmap 7-year power spectrum data at low
  multipoles: is it cosmological or systematics?
\newblock {\em Physical Review D}, 82(6):063002.

\bibitem[Koekemoer et~al., 2011]{koekemoer2011candels}
Koekemoer, a.~M., Faber, S., Ferguson, H.~C., Grogin, N.~A., Kocevski, D.~D.,
  Koo, D.~C., Lai, K., Lotz, J.~M., Lucas, R.~A., McGrath, E.~J., et~al.
  (2011).
\newblock Candels: The cosmic assembly near-infrared deep extragalactic legacy
  survey—the hubble space telescope observations, imaging data products, and
  mosaics.
\newblock {\em Astrophysical Journal Supplement Series}, 197(2):36.

\bibitem[Kraljic et~al., 2020]{kraljic2020and}
Kraljic, K., Dav{\'e}, R., and Pichon, C. (2020).
\newblock and yet it flips: connecting galactic spin and the cosmic web.
\newblock {\em Monthly Notices of the Royal Astronomical Society},
  493(1):362--381.

\bibitem[Kraljic et~al., 2021]{kraljic2021sdss}
Kraljic, K., Duckworth, C., Tojeiro, R., Alam, S., Bizyaev, D., Weijmans,
  a.-M., Boardman, N.~F., and Lane, R.~R. (2021).
\newblock Sdss-iv manga: 3d spin alignment of spiral and s0 galaxies.
\newblock {\em Monthly Notices of the Royal Astronomical Society}.

\bibitem[Krishnan et~al., 2021]{krishnan2021hints}
Krishnan, C., Mohayaee, R., Colg{\'a}in, E.~{\'O}., Sheikh-Jabbari, M., and
  Yin, L. (2021).
\newblock Hints of flrw breakdown from supernovae.
\newblock {\em arXiv preprint arXiv:2106.02532}.

\bibitem[Land and Magueijo, 2005]{land2005examination}
Land, K. and Magueijo, J. (2005).
\newblock Examination of evidence for a preferred axis in the cosmic radiation
  anisotropy.
\newblock {\em Physics Review Letters}, 95(7):071301.

\bibitem[Land et~al., 2008]{land2008galaxy}
Land, K., Slosar, a., Lintott, C., andreescu, D., Bamford, S., Murray, P.,
  Nichol, R., Raddick, M.~J., Schawinski, K., Szalay, A., et~al. (2008).
\newblock Galaxy zoo: the large-scale spin statistics of spiral galaxies in the
  sloan digital sky survey.
\newblock {\em Monthly Notices of the Royal Astronomical Society},
  388(4):1686--1692.

\bibitem[Le~Corre, 2021]{le2021recent}
Le~Corre, S. (2021).
\newblock Recent cosmological anisotropy explained by dark energy as universes
  of negative gravitational mass.
\newblock {\em Open Access Library Journal}, 8(6):1--12.

\bibitem[Lee et~al., 2019a]{lee2019galaxy}
Lee, J.~H., Pak, M., Lee, H.-R., and Song, H. (2019a).
\newblock Galaxy rotation coherent with the motions of neighbors: Discovery of
  observational evidence.
\newblock {\em Astrophysical Journal}, 872(1):78.

\bibitem[Lee et~al., 2019b]{lee2019mysterious}
Lee, J.~H., Pak, M., Song, H., Lee, H.-R., Kim, S., and Jeong, H. (2019b).
\newblock Mysterious coherence in several-megaparsec scales between galaxy
  rotation and neighbor motion.
\newblock {\em Astrophysical Journal}, 884(2):104.

\bibitem[Libeskind et~al., 2013]{libeskind2013velocity}
Libeskind, N.~I., Hoffman, Y., Forero-Romero, J., Gottl{\"o}ber, S., Knebe, A.,
  Steinmetz, M., and Klypin, a. (2013).
\newblock The velocity shear tensor: tracer of halo alignment.
\newblock {\em Monthly Notices of the Royal Astronomical Society},
  428(3):2489--2499.

\bibitem[Libeskind et~al., 2014]{libeskind2014universal}
Libeskind, N.~I., Knebe, A., Hoffman, Y., and Gottl{\"o}ber, S. (2014).
\newblock The universal nature of subhalo accretion.
\newblock {\em Monthly Notices of the Royal Astronomical Society},
  443(2):1274--1280.

\bibitem[Lietzen et~al., 2016]{lietzen2016discovery}
Lietzen, H., Tempel, E., Liivam{\"a}gi, L., Montero-Dorta, A., Einasto, M.,
  Streblyanska, A., Maraston, C., Rubi{\~n}o-Mart{\'\i}n, J., and Saar, E.
  (2016).
\newblock Discovery of a massive supercluster system at z\~{} 0.47.
\newblock {\em Astronomy and Astrophysics}, 588:L4.

\bibitem[Lin et~al., 2016]{lin2016significance}
Lin, H.-N., Li, X., and Chang, Z. (2016).
\newblock The significance of anisotropic signals hiding in the type ia
  supernovae.
\newblock {\em Monthly Notices of the Royal Astronomical Society},
  460(1):617--626.

\bibitem[Longo, 2011]{longo2011detection}
Longo, M.~J. (2011).
\newblock Detection of a dipole in the handedness of spiral galaxies with
  redshifts z~ 0.04.
\newblock {\em Physics Letters B}, 699(4):224--229.

\bibitem[Luongo et~al., 2021]{luongo2021larger}
Luongo, O., Muccino, M., Colg{\'a}in, E.~{\'O}., Sheikh-Jabbari, M., and Yin,
  L. (2021).
\newblock On larger $ h\_0 $ values in the cmb dipole direction.
\newblock {\em arXiv:2108.13228}.

\bibitem[MacGillivray and Dodd, 1985]{macgillivray1985anisotropy}
MacGillivray, H. and Dodd, R. (1985).
\newblock The anisotropy of the spatial orientations of galaxies in the local
  supercluster.
\newblock {\em Astronomy and Astrophysics}, 145:269--274.

\bibitem[Mariano and Perivolaropoulos, 2013]{mariano2013cmb}
Mariano, a. and Perivolaropoulos, L. (2013).
\newblock Cmb maximum temperature asymmetry axis: Alignment with other cosmic
  asymmetries.
\newblock {\em Physical Review D}, 87(4):043511.

\bibitem[McClintock et~al., 2006]{mcclintock2006spin}
McClintock, J.~E., Shafee, R., Narayan, R., Remillard, R.~A., Davis, S.~W., and
  Li, L.-X. (2006).
\newblock The spin of the near-extreme kerr black hole grs 1915+ 105.
\newblock {\em Astrophysical Journal}, 652(1):518.

\bibitem[M{\'e}sz{\'a}ros, 2019]{meszaros2019oppositeness}
M{\'e}sz{\'a}ros, A. (2019).
\newblock an oppositeness in the cosmology: Distribution of the gamma ray
  bursts and the cosmological principle.
\newblock {\em Astronomische Nachrichten}, 340(7):564--569.

\bibitem[Migkas et~al., 2021]{migkas2021cosmological}
Migkas, K., Pacaud, F., Schellenberger, G., Erler, J., Nguyen-Dang, N.,
  Reiprich, T., Ramos-Ceja, M., and Lovisari, L. (2021).
\newblock Cosmological implications of the anisotropy of ten galaxy cluster
  scaling relations.
\newblock {\em Astronomy and Astrophysics}, 649:A151.

\bibitem[Migkas et~al., 2020]{migkas2020probing}
Migkas, K., Schellenberger, G., Reiprich, T., Pacaud, F., Ramos-Ceja, M., and
  Lovisari, L. (2020).
\newblock Probing cosmic isotropy with a new x-ray galaxy cluster sample
  through the lx--t scaling relation.
\newblock {\em Astronomy and Astrophysics}, 636:A15.

\bibitem[Morh{\'a}{\v{c}} et~al., 2000]{morhavc2000identification}
Morh{\'a}{\v{c}}, M., Kliman, J., Matou{\v{s}}ek, V., Veselsk{\'y}, M., and
  Turzo, I. (2000).
\newblock Identification of peaks in multidimensional coincidence $\gamma$-ray
  spectra.
\newblock {\em Nuclear Instruments and Methods in Physics Research Section A:
  Accelerators, Spectrometers, Detectors and Associated Equipment},
  443(1):108--125.

\bibitem[Moster et~al., 2011]{moster2011cosmic}
Moster, B.~P., Somerville, R.~S., Newman, J.~A., and Rix, H.-W. (2011).
\newblock A cosmic variance cookbook.
\newblock {\em Astrophysical Journal}, 731(2):113.

\bibitem[Motloch et~al., 2021]{motloch2021observed}
Motloch, P., Yu, H.-R., Pen, U.-L., and Xie, Y. (2021).
\newblock an observed correlation between galaxy spins and initial conditions.
\newblock {\em Nature Astronomy}, 5(3):283--288.

\bibitem[Mudambi et~al., 2020]{mudambi2020estimation}
Mudambi, S.~P., Rao, A., Gudennavar, S., Misra, R., and Bubbly, S. (2020).
\newblock Estimation of the black hole spin in lmc x-1 using astrosat.
\newblock {\em Monthly Notices of the Royal Astronomical Society},
  498(3):4404--4410.

\bibitem[Myung, 2005]{myung2005holographic}
Myung, Y.~S. (2005).
\newblock Holographic principle and dark energy.
\newblock {\em Physics Letters B}, 610(1-2):18--22.

\bibitem[Nair and Abraham, 2010]{nair2010catalog}
Nair, P.~B. and Abraham, R.~G. (2010).
\newblock A catalog of detailed visual morphological classifications for 14,034
  galaxies in the sloan digital sky survey.
\newblock {\em Astrophysical Journal Supplement Series}, 186(2):427.

\bibitem[Ozsv{\'a}th and Sch{\"u}cking, 1962]{ozsvath1962finite}
Ozsv{\'a}th, I. and Sch{\"u}cking, E. (1962).
\newblock Finite rotating universe.
\newblock {\em Nature}, 193(4821):1168--1169.

\bibitem[Ozsvath and Sch{\"u}cking, 2001]{ozsvath2001approaches}
Ozsvath, I. and Sch{\"u}cking, E. (2001).
\newblock Approaches to g{\"o}del's rotating universe.
\newblock {\em Classical and Quantum Gravity}, 18(12):2243.

\bibitem[Panwar et~al., 2020]{panwar2020alignment}
Panwar, M., Sandhu, P.~K., Wadadekar, Y., and Jain, P.~J. (2020).
\newblock Alignment of radio galaxy axes using first catalogue.
\newblock {\em Monthly Notices of the Royal Astronomical Society},
  499(1):1226--1232.

\bibitem[Pathria, 1972]{pathria1972universe}
Pathria, R. (1972).
\newblock The universe as a black hole.
\newblock {\em Nature}, 240(5379):298--299.

\bibitem[Pecker, 1997]{pecker1997some}
Pecker, J.-C. (1997).
\newblock Some critiques of the big bang cosmology.
\newblock {\em Journal of Astrophysics and Astronomy}, 18(4):323--333.

\bibitem[P{\'e}rez-Gonz{\'a}lez et~al., 2015]{perez2015morphologies}
P{\'e}rez-Gonz{\'a}lez, P.~G., Mei, S., Shankar, F., Bernardi, M., Daddi, E.,
  Barro, G., Cabrera-Vives, G., Cattaneo, A., Dimauro, P., Gravet, R., et~al.
  (2015).
\newblock The morphologies of massive galaxies from z~ 3—witnessing the two
  channels of bulge growth.
\newblock {\em Astrophysical Journal}, 809(1):95.

\bibitem[Perivolaropoulos, 2014]{perivolaropoulos2014large}
Perivolaropoulos, L. (2014).
\newblock Large scale cosmological anomalies and inhomogeneous dark energy.
\newblock {\em Galaxies}, 2(1):22--61.

\bibitem[Piao, 2005]{piao2005possible}
Piao, Y.-S. (2005).
\newblock Possible explanation to a low cmb quadrupole.
\newblock {\em Physical Review D}, 71(8):087301.

\bibitem[Piao et~al., 2004]{piao2004suppressing}
Piao, Y.-S., Feng, B., and Zhang, X. (2004).
\newblock Suppressing the cmb quadrupole with a bounce from the contracting
  phase to inflation.
\newblock {\em Physical Review D}, 69(10):103520.

\bibitem[Pop{\l}awski, 2010a]{poplawski2010cosmology}
Pop{\l}awski, N.~J. (2010a).
\newblock Cosmology with torsion: an alternative to cosmic inflation.
\newblock {\em Physics Letters B}, 694(3):181--185.

\bibitem[Pop{\l}awski, 2010b]{poplawski2010radial}
Pop{\l}awski, N.~J. (2010b).
\newblock Radial motion into an einstein--rosen bridge.
\newblock {\em Physics Letters B}, 687(2-3):110--113.

\bibitem[Pourhasan et~al., 2014]{pourhasan2014out}
Pourhasan, R., Afshordi, N., and Mann, R.~B. (2014).
\newblock Out of the white hole: a holographic origin for the big bang.
\newblock {\em Journal of Cosmology and Astroparticle Physics}, 2014(04):005.

\bibitem[Rajpoot and Vacaru, 2017]{rajpoot2017supersymmetric}
Rajpoot, S. and Vacaru, S.~I. (2017).
\newblock On supersymmetric geometric flows and r2 inflation from scale
  invariant supergravity.
\newblock {\em Annals of Physics}, 384:20--60.

\bibitem[Ralston and Jain, 2004]{ralston2004virgo}
Ralston, J.~P. and Jain, P.~J. (2004).
\newblock The virgo alignment puzzle in propagation of radiation on
  cosmological scales.
\newblock {\em International Journal of Modern Physics D}, 13(09):1857--1877.

\bibitem[Reynolds, 2021]{reynolds2021observational}
Reynolds, C.~S. (2021).
\newblock Observational constraints on black hole spin.
\newblock {\em Annual Review of Astronomy and Astrophysics}, 59:117--154.

\bibitem[Rinaldi et~al., 2022]{rinaldi2022matrix}
Rinaldi, E., Han, X., Hassan, M., Feng, Y., Nori, F., McGuigan, M., and Hanada,
  M. (2022).
\newblock Matrix-model simulations using quantum computing, deep learning, and
  lattice monte carlo.
\newblock {\em Physical Review X Quantum}, 3(1):010324.

\bibitem[Rodrigues, 2008]{rodrigues2008anisotropic}
Rodrigues, D.~C. (2008).
\newblock anisotropic cosmological constant and the cmb quadrupole anomaly.
\newblock {\em Physical Review D}, 77(2):023534.

\bibitem[Sanders, 2003]{sanders2003missing}
Sanders, R. (2003).
\newblock Missing mass as evidence for modified newtonian dynamics at low
  accelerations.
\newblock {\em Modern Physics Letters A}, 18(27):1861--1875.

\bibitem[Santos et~al., 2015]{santos2015influence}
Santos, L., Cabella, P., Villela, T., and Zhao, W. (2015).
\newblock Influence of planck foreground masks in the large angular scale
  quadrant cmb asymmetry.
\newblock {\em Astronomy and Astrophysics}, 584:A115.

\bibitem[Schwarz et~al., 2004]{schwarz2004low}
Schwarz, D.~J., Starkman, G.~D., Huterer, D., and Copi, C.~J. (2004).
\newblock Is the low-l microwave background cosmic?
\newblock {\em Physics Review Letters}, 93(22):221301.

\bibitem[Secrest et~al., 2021]{secrest2021test}
Secrest, N.~J., von Hausegger, S., Rameez, M., Mohayaee, R., Sarkar, S., and
  Colin, J. (2021).
\newblock A test of the cosmological principle with quasars.
\newblock {\em Astrophysical Journal Letters}, 908(2):L51.

\bibitem[Semenaite et~al., 2021]{semenaite2021cosmological}
Semenaite, A., S{\'a}nchez, A.~G., Pezzotta, a., Hou, J., Scoccimarro, R.,
  Eggemeier, A., Crocce, M., Chuang, C.-H., Smith, A., Zhao, C., et~al. (2021).
\newblock Cosmological implications of the full shape of anisotropic clustering
  measurements in boss and eboss.
\newblock {\em arXiv:2111.03156}.

\bibitem[Seshavatharam, 2010]{seshavatharam2010physics}
Seshavatharam, U. (2010).
\newblock Physics of rotating and expanding black hole universe.
\newblock {\em Progress in Physics}, 2:7--14.

\bibitem[Seshavatharam and Lakshminarayana,
  2014]{seshavatharam2014understanding}
Seshavatharam, U. and Lakshminarayana, S. (2014).
\newblock Understanding black hole cosmology and the cosmic halt.
\newblock {\em Journal of Advanced Research in Astrophysics and Astronomy},
  1(1):1--27.

\bibitem[Seshavatharam and Lakshminarayana, 2020a]{seshavatharam2020integrated}
Seshavatharam, U. and Lakshminarayana, S. (2020a).
\newblock an integrated model of a light speed rotating universe.
\newblock {\em International Astronomy and Astrophysics Research Journal},
  pages 74--82.

\bibitem[Seshavatharam and Lakshminarayana, 2020b]{seshavatharam2020light}
Seshavatharam, U. and Lakshminarayana, S. (2020b).
\newblock Light speed expansion and rotation of a primordial black hole
  universe having internal acceleration.
\newblock {\em International Astronomy and Astrophysics Research Journal},
  pages 9--27.

\bibitem[Shamir, 2011]{shamir2011ganalyzer}
Shamir, L. (2011).
\newblock Ganalyzer: A tool for automatic galaxy image analysis.
\newblock {\em Astrophysical Journal}, 736(2):141.

\bibitem[Shamir, 2012]{shamir2012handedness}
Shamir, L. (2012).
\newblock Handedness asymmetry of spiral galaxies with z< 0.3 shows cosmic
  parity violation and a dipole axis.
\newblock {\em Physics Letters B}, 715(1-3):25--29.

\bibitem[Shamir, 2013]{shamir2013color}
Shamir, L. (2013).
\newblock Color differences between clockwise and counterclockwise spiral
  galaxies.
\newblock {\em Galaxies}, 1(3):210--215.

\bibitem[Shamir, 2016]{shamir2016asymmetry}
Shamir, L. (2016).
\newblock Asymmetry between galaxies with clockwise handedness and
  counterclockwise handedness.
\newblock {\em Astrophysical Journal}, 823(1):32.

\bibitem[Shamir, 2017a]{shamir2017colour}
Shamir, L. (2017a).
\newblock Colour asymmetry between galaxies with clockwise and counterclockwise
  handedness.
\newblock {\em Astrophysics and Space Science}, 362(2):33.

\bibitem[Shamir, 2017b]{shamir2017large}
Shamir, L. (2017b).
\newblock Large-scale photometric asymmetry in galaxy spin patterns.
\newblock {\em Publications of the Astronomical Society of Australia}, 34:e44.

\bibitem[Shamir, 2017c]{shamir2017photometric}
Shamir, L. (2017c).
\newblock Photometric asymmetry between clockwise and counterclockwise spiral
  galaxies in sdss.
\newblock {\em Publications of the Astronomical Society of Australia}, 34:e011.

\bibitem[Shamir, 2019]{shamir2019large}
Shamir, L. (2019).
\newblock Large-scale patterns of galaxy spin rotation show cosmological-scale
  parity violation and multipoles.
\newblock {\em arXiv}, page 1912.05429.

\bibitem[Shamir, 2020a]{shamir2020asymmetry}
Shamir, L. (2020a).
\newblock Asymmetry between galaxies with different spin patterns: A comparison
  between cosmos, sdss, and pan-starrs.
\newblock {\em Open Astronomy}, 29(1):15--27.

\bibitem[Shamir, 2020b]{shamir2020pasa}
Shamir, L. (2020b).
\newblock Galaxy spin direction distribution in hst and sdss show similar
  large-scale asymmetry.
\newblock {\em Publications of the Astronomical Society of Australia}, 37:e053.

\bibitem[Shamir, 2020c]{shamir2020large}
Shamir, L. (2020c).
\newblock Large-scale asymmetry between clockwise and counterclockwise galaxies
  revisited.
\newblock {\em Astronomische Nachrichten}, 341(3):324.

\bibitem[Shamir, 2020d]{shamir2020patterns}
Shamir, L. (2020d).
\newblock Patterns of galaxy spin directions in sdss and pan-starrs show parity
  violation and multipoles.
\newblock {\em Astrophysics and Space Science}, 365:136.

\bibitem[Shamir, 2021a]{shamir2021particles}
Shamir, L. (2021a).
\newblock analysis of the alignment of non-random patterns of spin directions
  in populations of spiral galaxies.
\newblock {\em Particles}, 4(1):11--28.

\bibitem[Shamir, 2021b]{shamir2021large}
Shamir, L. (2021b).
\newblock Large-scale asymmetry in galaxy spin directions: evidence from the
  southern hemisphere.
\newblock {\em Publications of the Astronomical Society of Australia}, 38:e037.

\bibitem[Shamir, 2022]{shamir2022new}
Shamir, L. (2022).
\newblock New evidence and analysis of cosmological-scale asymmetry in galaxy
  spin directions.
\newblock {\em Journal of Astrophysics and Astronomy}, In Press.

\bibitem[Shor et~al., 2021]{shor2021representation}
Shor, O., Benninger, F., and Khrennikov, a. (2021).
\newblock Representation of the universe as a dendrogramic hologram endowed
  with relational interpretation.
\newblock {\em Entropy}, 23(5):584.

\bibitem[Sivaram and Arun, 2012]{sivaram2012primordial}
Sivaram, C. and Arun, K. (2012).
\newblock Primordial rotation of the universe, hydrodynamics, vortices and
  angular momenta of celestial objects.
\newblock {\em Open Astronomy}, 5:7--11.

\bibitem[Sivaram and Arun, 2013]{sivaram2013holography}
Sivaram, C. and Arun, K. (2013).
\newblock Holography, dark energy and entropy of large cosmic structures.
\newblock {\em Astrophysics and Space Science}, 348(1):217--219.

\bibitem[Skeivalas et~al., 2021]{skeivalas2021predictive}
Skeivalas, J., Par{\v{s}}eli{\=u}nas, E., and {\v{S}}likas, D. (2021).
\newblock The predictive model for the universe rotation axis identification
  upon applying the solar system coordinate net in the milky way galaxy.
\newblock {\em Indian Journal of Physics}, pages 1--10.

\bibitem[Stuckey, 1994]{stuckey1994observable}
Stuckey, W. (1994).
\newblock The observable universe inside a black hole.
\newblock {\em American Journal of Physics}, 62(9):788--795.

\bibitem[Su and Chu, 2009]{su2009universe}
Su, S.-C. and Chu, M.-C. (2009).
\newblock Is the universe rotating?
\newblock {\em Astrophysical Journal}, 703(1):354.

\bibitem[Susskind, 1995]{susskind1995world}
Susskind, L. (1995).
\newblock The world as a hologram.
\newblock {\em Journal of Mathematical Physics}, 36(11):6377--6396.

\bibitem[Takahashi, 2004]{takahashi2004shapes}
Takahashi, R. (2004).
\newblock Shapes and positions of black hole shadows in accretion disks and
  spin parameters of black holes.
\newblock {\em Astrophysical Journal}, 611(2):996.

\bibitem[Tatum et~al., 2018a]{tatum2018flat}
Tatum, E.~T. et~al. (2018a).
\newblock Why flat space cosmology is superior to standard inflationary
  cosmology.
\newblock {\em Journal of Modern Physics}, 9(10):1867.

\bibitem[Tatum et~al., 2018b]{tatum2018clues}
Tatum, E.~T., Seshavatharam, U., et~al. (2018b).
\newblock Clues to the fundamental nature of gravity, dark energy and dark
  matter.
\newblock {\em Journal of Modern Physics}, 9(08):1469.

\bibitem[Taylor and Jagannathan, 2016]{taylor2016alignments}
Taylor, A. and Jagannathan, P. (2016).
\newblock Alignments of radio galaxies in deep radio imaging of elais n1.
\newblock {\em Monthly Notices of the Royal Astronomical Society},
  459(1):L36--L40.

\bibitem[Tempel and Libeskind, 2013]{tempel2013galaxy}
Tempel, E. and Libeskind, N.~I. (2013).
\newblock Galaxy spin alignment in filaments and sheets: observational
  evidence.
\newblock {\em Astrophysical Journal Letters}, 775(2):L42.

\bibitem[Tempel et~al., 2014]{tempel2014detecting}
Tempel, E., Stoica, R., Martinez, V.~J., Liivam{\"a}gi, L., Castellan, G., and
  Saar, E. (2014).
\newblock Detecting filamentary pattern in the cosmic web: a catalogue of
  filaments for the sdss.
\newblock {\em Monthly Notices of the Royal Astronomical Society},
  438(4):3465--3482.

\bibitem[Tempel et~al., 2013]{tempel2013evidence}
Tempel, E., Stoica, R.~S., and Saar, E. (2013).
\newblock Evidence for spin alignment of spiral and elliptical/s0 galaxies in
  filaments.
\newblock {\em Monthly Notices of the Royal Astronomical Society},
  428(2):1827--1836.

\bibitem[Tiwari and Jain, 2015]{tiwari2015dipole}
Tiwari, P. and Jain, P.~J. (2015).
\newblock Dipole anisotropy in integrated linearly polarized flux density in
  nvss data.
\newblock {\em Monthly Notices of the Royal Astronomical Society},
  447(3):2658--2670.

\bibitem[Tiwari and Nusser, 2016]{tiwari2016revisiting}
Tiwari, P. and Nusser, A. (2016).
\newblock Revisiting the nvss number count dipole.
\newblock {\em Journal of Cosmology and Astroparticle Physics}, 2016(03):062.

\bibitem[Velten and Gomes, 2020]{velten2020hubble}
Velten, H. and Gomes, S. (2020).
\newblock Is the hubble diagram of quasars in tension with concordance
  cosmology?
\newblock {\em Physical Review D}, 101(4):043502.

\bibitem[Volonteri et~al., 2005]{volonteri2005distribution}
Volonteri, M., Madau, P., Quataert, E., and Rees, M.~J. (2005).
\newblock The distribution and cosmic evolution of massive black hole spins.
\newblock {\em Astrophysical Journal}, 620(1):69.

\bibitem[Wang et~al., 2018]{wang2018spin}
Wang, P., Guo, Q., Kang, X., and Libeskind, N.~I. (2018).
\newblock The spin alignment of galaxies with the large-scale tidal field in
  hydrodynamic simulations.
\newblock {\em Astrophysical Journal}, 866(2):138.

\bibitem[Wang and Kang, 2017]{wang2017general}
Wang, P. and Kang, X. (2017).
\newblock A general explanation on the correlation of dark matter halo spin
  with the large-scale environment.
\newblock {\em Monthly Notices of the Royal Astronomical Society Letters},
  468(1):L123--L127.

\bibitem[Wang and Kang, 2018]{wang2018build}
Wang, P. and Kang, X. (2018).
\newblock The build up of the correlation between halo spin and the large-scale
  structure.
\newblock {\em Monthly Notices of the Royal Astronomical Society Letters},
  473(2):1562--1569.

\bibitem[Webb et~al., 2011]{webb2011indications}
Webb, J., King, J., Murphy, M., Flambaum, V., Carswell, R., and Bainbridge, M.
  (2011).
\newblock Indications of a spatial variation of the fine structure constant.
\newblock {\em Physics Review Letters}, 107(19):191101.

\bibitem[Weeks et~al., 2004]{weeks2004well}
Weeks, J., Luminet, J.-P., Riazuelo, A., and Lehoucq, R. (2004).
\newblock Well-proportioned universes suppress the cosmic microwave background
  quadrupole.
\newblock {\em Monthly Notices of the Royal Astronomical Society},
  352(1):258--262.

\bibitem[Yeung and Chu, 2022]{yeung2022directional}
Yeung, S. and Chu, M.-C. (2022).
\newblock Directional variations of cosmological parameters from the planck cmb
  data.
\newblock {\em Physical Review D}.

\bibitem[Zhang et~al., 2009]{zhang2009spin}
Zhang, Y., Yang, X., Faltenbacher, a., Springel, V., Lin, W., and Wang, H.
  (2009).
\newblock The spin and orientation of dark matter halos within cosmic
  filaments.
\newblock {\em Astrophysical Journal}, 706(1):747.

\bibitem[Zhao and Xia, 2021]{zhao2021tomographic}
Zhao, D. and Xia, J.-Q. (2021).
\newblock A tomographic test of cosmic anisotropy with the recently-released
  quasar sample.
\newblock {\em The European Physical Journal C}, 81(10):1--11.

\bibitem[Zhe et~al., 2015]{zhe2015quadrupole}
Zhe, C., Xin, L., and Sai, W. (2015).
\newblock Quadrupole-octopole alignment of cmb related to the primordial power
  spectrum with dipolar modulation in anisotropic spacetime.
\newblock {\em Chinese Physics C}, 39(5):055101.

\bibitem[Zhou et~al., 2021]{zhou2021clustering}
Zhou, R., Newman, J.~A., Mao, Y.-Y., Meisner, A., Moustakas, J., Myers, A.~D.,
  Prakash, A., Zentner, a.~R., Brooks, D., Duan, Y., et~al. (2021).
\newblock The clustering of desi-like luminous red galaxies using photometric
  redshifts.
\newblock {\em Monthly Notices of the Royal Astronomical Society},
  501(3):3309--3331.

\end{thebibliography}

\end{document}